\documentclass[useAMS,usenatbib]{mn2e}

\usepackage{natbib} %for references
\usepackage{multirow} %for multicolumns in tables
\usepackage[caption=false]{subfig} %for figures over multiple pages
\usepackage{enumerate} %for lists
\usepackage[T1]{fontenc} %for emphasis in titles
\usepackage{amsmath} %for equations
\usepackage{amssymb} %for equations
\usepackage{rotating} %for sideways figures
\usepackage{chngcntr} %for appendix figure numbers
\usepackage{appendix} %for appendices
\usepackage{book tabs} %for horizontal lines in tables

\def\reff@jnl#1{{\rm#1\/}}

\def\aj{\reff@jnl{AJ}}                  % Astronomical Journal
\def\araa{\reff@jnl{ARA\&A}}            % Annual Review of Astron and Astrophys
\def\apj{\reff@jnl{ApJ}}                % Astrophysical Journal
\def\apjl{\reff@jnl{ApJ}}               % Astrophysical Journal, Letters
\def\apjs{\reff@jnl{ApJS}}              % Astrophysical Journal, Supplement
\def\ao{\reff@jnl{Appl.Optics}}         % Applied Optics
\def\apss{\reff@jnl{Ap\&SS}}            % Astrophysics and Space Science
\def\aap{\reff@jnl{A\&A}}               % Astronomy and Astrophysics
\def\aapr{\reff@jnl{A\&A~Rev.}}         % Astronomy and Astrophysics Reviews
\def\aaps{\reff@jnl{A\&AS}}             % Astronomy and Astrophysics, Supplement
\def\azh{\reff@jnl{AZh}}                        % Astronomicheskii Zhurnal
\def\baas{\reff@jnl{BAAS}}              % Bulletin of the AAS
\def\jrasc{\reff@jnl{JRASC}}            % Journal of the RAS of Canada
\def\memras{\reff@jnl{MmRAS}}           % Memoirs of the RAS
\def\mnras{\reff@jnl{MNRAS}}            % Monthly Notices of the RAS
\def\pra{\reff@jnl{Phys.Rev.A}}         % Physical Review A: General Physics
\def\prb{\reff@jnl{Phys.Rev.B}}         % Physical Review B: Solid State
\def\prc{\reff@jnl{Phys.Rev.C}}         % Physical Review C
\def\prd{\reff@jnl{Phys.Rev.D}}         % Physical Review D
\def\prl{\reff@jnl{Phys.Rev.Lett}}      % Physical Review Letters
\def\pasp{\reff@jnl{PASP}}              % Publications of the ASP
\def\pasj{\reff@jnl{PASJ}}              % Publications of the ASJ
\def\qjras{\reff@jnl{QJRAS}}            % Quarterly Journal of the RAS
\def\skytel{\reff@jnl{S\&T}}            % Sky and Telescope
\def\solphys{\reff@jnl{Solar~Phys.}}    % Solar Physics
\def\sovast{\reff@jnl{Soviet~Ast.}}     % Soviet Astronomy
\def\ssr{\reff@jnl{Space~Sci.Rev.}}     % Space Science Reviews
\def\zap{\reff@jnl{ZAp}}                        % Zeitschrift fuer Astrophysik
\def\nat{\reff@jnl{Nature}}             % Nature 

\title[Constraining Very Small Grain Abundances In Galactic Cold Cores]{Using cm Observations to Constrain the Abundance of Very Small Dust Grains in Galactic Cold Cores}

\author[Tibbs et al.]{C.T.~Tibbs$^{1,2}$\thanks{ESA Research Fellow}\thanks{E-mail: ctibbs@cosmos.esa.int}, R.~Paladini$^{2}$, K.~Cleary$^{3}$, S.J.C.~Muchovej$^{3}$, A.M.M.~Scaife$^{4}$, \and M.A.~Stevenson$^{3}$, R.J.~Laureijs$^{1}$, N.~Ysard$^{5}$, K.J.B.~Grainge$^{4}$, Y.C.~Perrott$^{6}$, \and C.~Rumsey$^{6}$, J.~Villadsen$^{3}$ \\
$^{1}$Scientific Support Office, Directorate of Science and Robotic Exploration, European Space Research and Technology Centre (ESA/ESTEC), \\ Keplerlaan 1, 2201 AZ, Noordwijk, The Netherlands \\
$^{2}$Infrared Processing Analysis Center, California Institute of Technology, Pasadena, CA 91125, USA \\
$^{3}$Cahill Center for Astronomy and Astrophysics, California Institute of Technology, Pasadena, CA 91125, USA \\
$^{4}$Jodrell Bank Centre for Astrophysics, The University of Manchester, Manchester, M13 9PL, UK \\
$^{5}$IAS, Universit\'{e} Paris-Sud, 91405 Orsay cedex, France \\
$^{6}$Astrophysics Group, Cavendish Laboratory, University of Cambridge, Cambridge, CB3 0HE, UK }

\begin{document}

\date{Accepted 2015 November 20.  Received 2015 November 19; in original form 2015 August 13.}

\pagerange{\pageref{firstpage}--\pageref{lastpage}} \pubyear{}

\maketitle

\label{firstpage}

%%%%%%% Abstract %%%%%%%%%%%%%%%%%%%%%%%%%%%%%%%%%%%%%%%%%%%%

\begin{abstract}
In this analysis we illustrate how the relatively new emission mechanism known as spinning dust can be used to characterize dust grains in the interstellar medium. We demonstrate this by using spinning dust emission observations to constrain the abundance of very small dust grains~($a$~$\lesssim$~10~nm) in a sample of Galactic cold cores. Using the physical properties of the cores in our sample as inputs to a spinning dust model, we predict the expected level of emission at a wavelength of 1~cm for four different very small dust grain abundances, which we constrain by comparing to 1~cm CARMA observations. For all of our cores we find a depletion of very small grains, which we suggest is due to the process of grain growth. This work represents the first time that spinning dust emission has been used to constrain the physical properties of interstellar dust grains.
\end{abstract}

%%%%%%% Key Words %%%%%%%%%%%%%%%%%%%%%%%%%%%%%%%%%%%%%%%%%%%%


%%%%%%% Introduction %%%%%%%%%%%%%%%%%%%%%%%%%%%%%%%%%%%%%%%%%

\section{Introduction}
\label{Sec:Intro}

Interstellar dust is a key component of the cycle of matter in the interstellar medium~(ISM). Although, our understanding of the ISM has improved greatly over the last 25~years, there is still much uncertainty regarding the physical and evolutionary properties of interstellar dust grains~\citep[e.g., see][for a complete review]{Draine:03}. Dust grains are heated by stellar photons and re-radiate this absorbed energy as emission at infrared~(IR) wavelengths. The emission observed at mid- to far-IR wavelengths is a combination of thermal equilibrium emission, produced by a population of big grains~($a$~$\sim$10~--~1000~nm), and stochastic emission, produced by a population of very small grains/large molecules~($a$~$\lesssim$~10~nm). Throughout this analysis we refer to this population of very small grains/large molecules simply as very small grains.

Although it is recognized that dust grains consist mainly of silicates and carbon, the nature of their exact shape, abundance, and size distribution is still unclear. The observed polarisation of starlight~\citep[]{Hall:49, Hiltner:49} implies that dust grains are aligned with the magnetic field and that they are not spherically symmetric, but the actual shape is still unknown. Abundance estimates are constrained by measuring the depletion of elements from the gas phase, but there is still some debate over the exact values~\citep[e.g.,][]{Jenkins:09}. The size distribution of big grains can be inferred from observations of the wavelength-dependent interstellar extinction, but this is not possible for very small grains since they are in the Rayleigh regime, and the extinction they produce is not sensitive to their size distribution but only to their total volume i.e., the extinction curve only allows us to infer the presence of very small grains but provides no constraint on their size distribution~\citep[][]{Kim:94}.

Additionally, evolutionary properties such as how dust grains evolve from diffuse to dense media are still uncertain, particularly the process of grain growth. Since dust grains in the ISM are the seeds from which proto-planetary systems around young stellar objects are formed, one of the outstanding questions in the formation process of planets is how to account for the gap between the sub-micron sized grains found in the diffuse ISM and the millimetre sized grains required to form proto-planets, emphasizing the importance of the grain growth process.

To date, all of the information obtained about the properties of interstellar dust grains has been gained from observations at ultraviolet/optical/IR wavelengths based on interstellar extinction, scattering, absorption, and depletion measurements, along with emission and polarimetry observations. In this work, we demonstrate how observations at cm wavelengths can also be used to help characterise the properties of interstellar dust grains. To do this we use the recently discovered emission mechanism known as spinning dust emission: emission from very small dust grains characterised by an electric dipole moment, which when spinning, produce electric dipole radiation~\citep{DaL:98}. The spinning dust theory states that the very small grains that are emitting stochastically at mid-IR wavelengths are the same very small grains that are producing the spinning dust emission, allowing us to use this mechanism to investigate the physics of this population of very small dust grains. 

Spinning dust emission has been detected in a variety of Galactic environments such as molecular clouds~\citep{Casassus:08, Planck_Dickinson:11, Tibbs:10, Tibbs:13a, Genova-Santos:15}, dark clouds~\citep{Casassus:06, Scaife:09, Dickinson:10}, H\textsc{ii} regions~\citep{Dickinson:07, Tibbs:12a, Battistelli:15}, and reflection nebulae~\citep{Castellanos:11, Genova-Santos:11}, however, it has never been used to constrain the physical properties of interstellar dust grains. Complex spinning dust models now exist, incorporating a variety of excitation and damping mechanisms acting on the dust grains~\citep[e.g.,][]{Ali-Haimoud:09, Hoang:10, Hoang:11, Silsbee:11}. In the present analysis we show that, by taking advantage of these sophisticated models, spinning dust emission can be used as a probe of the very small dust grain abundance. 

Star formation is the result of gravitational instability occurring in cold, dense structures known as pre-stellar cores~\citep{Bergin:07}. These dense environments also play a vital role in the life cycle of dust in the ISM, and represent an ideal location in which to study dust grain evolution.~\citet{Tibbs:15a} were the first to search for spinning dust emission in Galactic cores using 1~cm observations obtained with the CARMA interferometer. By comparing these observations to the expected level of spinning dust emission, they concluded that they could not completely rule out spinning dust emission originating from these dense environments. However, they found that, if spinning dust emission is present in cores, it is at a lower level than the one predicted by their modelling analysis.~\citet{Tibbs:15a} performed their modelling by combining parameters constrained through ancillary data (e.g., from far-IR and CO observations) with others set to canonical values. This approach already represents a major step forward with respect to previously published analyses where only canonical parameter values were used. In this paper, we expand upon the \citet{Tibbs:15a} modelling analysis and show that not only is it possible to successfully reproduce the observed CARMA cm emission, but we can also constrain the abundance of very small grains in these cores.

This paper is organized as follows. In Section~\ref{Sec:Source_Selection} we describe the sample of sources used in this analysis, while in Section~\ref{Sec:Spinning_Dust} we discuss how we model the spinning dust emission. In Section~\ref{sec:Abundance_Coagulation} we compare the predicted level of spinning dust emission with our observations, which allows us to constrain the abundance of very small grains. We also discuss the implications for the dust grain evolution in these dense environments. Finally, in Section~\ref{Sec:Conclusions} we present our conclusions.

\begin{table*}
\begin{center}
\caption{Summary of the 15 cold clumps used in this analysis.} %RA and DEC from the Planck ECC Catalogue
\begin{tabular}{cccccc}
\hline
Target & R.A. & Decl. & Distance & $T_{\mathrm{gas}}$  \\
 & (J2000) & (J2000) & (kpc) & (K) \\
\hline
\hline

ECC181 G102.19+15.24		& 20:41:10.74 		& +67:21:44.3 		& 0.33$^{a}$ 	& 9.6	 \\
ECC189 G103.71+14.88 		& 20:53:30.29 		& +68:19:32.9 		& 0.29$^{a}$ 	& 9.7 \\
ECC190 G103.77+13.90 		& 21:02:09.19 		& +67:45:51.8 		& 0.29$^{a}$ 	& 11.1 \\
ECC191 G103.90+13.97 		& 21:02:23.24 		& +67:54:43.3 		& 0.29$^{a}$ 	& 11.1 \\
ECC223 G113.42+16.97 		& 21:59:59.03 		& +76:34:08.7 		& 0.99$^{b}$ 	& 8.9 \\
ECC224 G113.62+15.01 		& 22:21:37.34 		& +75:06:33.5 		& 0.86$^{b}$ 	& 8.0	 \\
ECC225 G113.75+14.90 		& 22:24:16.23 		& +75:05:01.8 		& 0.88$^{b}$ 	& 8.6 \\
ECC229 G114.67+14.47 		& 22:39:35.57 		& +75:11:34.0 		& 0.77$^{b}$ 	& 10.3 \\
ECC276 G127.88+02.66 		& 01:38:39.14 		& +65:05:06.5 		& 1.16$^{b}$ 	& 12.6 \\
ECC332 G149.41+03.37 		& 04:17:09.10 		& +55:17:39.4 		& 0.18$^{b}$ 	& 8.7 \\
ECC334 G149.58+03.45 		& 04:18:23.96 		& +55:13:30.6 		& 0.20$^{b}$ 	& 8.7 \\
ECC335 G149.65+03.54 		& 04:19:11.28 		& +55:14:44.4 		& 0.17$^{b}$ 	& 8.1 \\
ECC340 G151.45+03.95 		& 04:29:56.29 		& +54:14:51.7 		& 0.19$^{a}$ 	& 10.1 \\
ECC345 G154.07+05.09 		& 04:47:23.41 		& +53:03:31.4 		& 0.34$^{b}$ 	& 7.4 \\
ECC346 G154.07+05.21 		& 04:47:57.83 		& +53:07:51.2 		& 0.23$^{b}$ 	& 10.0 \\

\hline
\label{Table:Sources}
\end{tabular}
\end{center}
\vspace{-0.6cm}
\textbf{Notes}: $^{a}$Distances are based on association with known Lynds dark nebulae~\citep{Hilton:95}. $^{b}$Distances are kinematic distances based on $^{13}$CO observations~\citep{Wu:12}. $T_{\mathrm{gas}}$ values taken from \citet{Wu:12}.
\end{table*}

%%%%%%% Source Selection %%%%%%%%%%%%%%%%%%%%%%%%%%%%%%%%%%%%%%%%%

\section{Source Selection}
\label{Sec:Source_Selection}

In this analysis we use the sample defined by~\citet{Tibbs:15a}. This is a sample of 15 sources~(listed in Table~\ref{Table:Sources}) selected from the \textit{Planck} Early Cold Core~(ECC) catalogue, which was released as part of the \textit{Planck} Early Release Compact Source Catalogue~\citep{Planck_Chary:11}. As described by~\citet{Tibbs:15a}, the sample of 15 clumps\footnote{Since the angular resolution of \textit{Planck} is~$\sim$5~arcmin in the far-IR/sub-mm bands, it can not detect individual cores, and therefore we refer to the \textit{Planck} sources as clumps.} was observed at 1~cm with the CARMA interferometer, and using the available \textit{Herschel Space Observatory} data, a total of 34 cores were identified, whose physical properties~(e.g., mean column density, $\bar{N}$$_{\mathrm{H}}$, mean dust temperature, $\bar{T}$$_{\mathrm{d}}$, mean density, $\bar{n}$$_{\mathrm{H}}$, mean radiation field, $\bar{G}$$_{\mathrm{0}}$, mass and size) were estimated~--~for full details see~\citet[][]{Tibbs:15a}. In this work we use the observed CARMA 1~cm flux densities and the physical properties of the cores derived from the \textit{Herschel} data. For convenience, and to ensure that this paper is self-contained, we have listed all of the properties that we use in this analysis in~Table~\ref{Table:Clump_Props}~(see first 8 columns).

\begin{table*}
\begin{center}
\caption{Physical properties of the cores, including the core size, mean density ($\bar{n}$$_{\mathrm{H}}$), mean radiation field ($\bar{G}$$_{\mathrm{0}}$), observed 1~cm flux density ($S_{1~\mathrm{cm}}^{\mathrm{observed}}$), and the very small dust grain abundance~($b_{\mathrm{C}}$) that is constrained by this analysis.}
\begin{tabular}{ccccccccc}
\hline
Clump & Core & R.A. & Decl. & Size & $\bar{n}$$_{\mathrm{H}}$ & $\bar{G}$$_{\mathrm{0}}$ & $S_{\mathrm{1~cm}}^{\mathrm{observed}}$ & $b_{\mathrm{C}}$ \\
 & & (J2000) & (J2000) & (pc) & ($10^{3}~\mathrm{H~cm}^{-3}$) & & (mJy) & \\
\hline
\hline

%Spinning Dust Parameters:
%     n_H (H/cm^3): see below
%     T_gas (K): see below
%     G0 : see below
%     x_H : 0.0000000
%     x_C : 1.0000000e-06
%     y : 0.99900000
%     gamma : 0.0000000
%     mu : 9.3000000
%     Line : 16

\multirow{3}{*}{ECC181} 	& 1 & 20:41:13.2 & 67:20:34.2 & 0.23 & 25.6 $\pm$ 3.4 & 0.086 $\pm$ 0.008 & <~2.87 & <~1$\times$10$^{-5}$ \\
 					& 2 & 20:40:56.2 & 67:22:54.4 & 0.18 & 32.0 $\pm$ 4.2 & 0.090 $\pm$ 0.009 & <~1.66 & <~1$\times$10$^{-5}$ \\
 					& 3 & 20:40:31.9 & 67:20:48.1 & 0.13 & 41.3 $\pm$ 5.5 & 0.078 $\pm$ 0.007 & <~0.86 & <~1$\times$10$^{-5}$ \\
\\
\multirow{1}{*}{ECC189} 	& 1 & 20:53:35.4 & 68:19:19.9 & 0.17 & 18.7 $\pm$ 2.5 & 0.120 $\pm$ 0.012 & <~1.30 & <~1$\times$10$^{-5}$ \\
\\
\multirow{2}{*}{ECC190} 	& 1 & 21:01:54.4 & 67:43:45.7 & 0.20 & 31.6 $\pm$ 4.2 & 0.122 $\pm$ 0.012 & <~2.15 & <~1$\times$10$^{-5}$ \\
 					& 2 & 21:02:21.5 & 67:45:37.7 & 0.21 & 25.4 $\pm$ 3.4 & 0.146 $\pm$ 0.015 & <~2.48 & <~1$\times$10$^{-5}$ \\
\\
\multirow{1}{*}{ECC191} 	& 1 & 21:02:23.2 & 67:54:15.2 & 0.21 & 63.8 $\pm$ 8.4 & 0.083 $\pm$ 0.008 & <~4.80 & <~1$\times$10$^{-5}$ \\
\\
\multirow{1}{*}{ECC223} 	& 1 & 21:59:38.8 & 76:33:13.1 & 1.06 & 5.3 $\pm$ 0.7 & 0.108 $\pm$ 0.011 & <~5.69 & <~1$\times$10$^{-5}$ \\
\\
\multirow{3}{*}{ECC224} 	& 1 & 22:21:27.6 & 75:04:26.0 & 0.48 & 13.4 $\pm$ 1.8 & 0.112 $\pm$ 0.011 & <~2.36 & <~1$\times$10$^{-5}$ \\
 					& 2 & 22:21:41.2 & 75:06:06.0 & 0.51 & 11.2 $\pm$ 1.5 & 0.129 $\pm$ 0.013 & <~2.72 & <~1$\times$10$^{-5}$ \\
 					& 3 & 22:21:50.5 & 75:09:09.5 & 0.28 & 14.6 $\pm$ 1.9 & 0.138 $\pm$ 0.014 & <~0.81 & $\le$~1$\times$10$^{-5}$ \\
\\
\multirow{1}{*}{ECC225} 	& 1 & 22:24:09.1 & 75:04:33.1 & 0.71 & 7.5 $\pm$ 1.0 & 0.105 $\pm$ 0.010 & <~4.49 & <~1$\times$10$^{-5}$ \\
\\
\multirow{4}{*}{ECC229} 	& 1 & 22:39:39.5 & 75:12:01.5 & 0.34 & 97.0 $\pm$ 12.7 & 0.040 $\pm$ 0.003 & 2.87~$\pm$~0.88 & 0$\times$10$^{-5}$ \\
 					& 2 & 22:38:47.9 & 75:11:27.2 & 0.24 & 123.7 $\pm$ 17.3 & 0.298 $\pm$ 0.041 & 4.12~$\pm$~0.91 & 1$\times$10$^{-5}$ \\
 					& 3 & 22:39:31.6 & 75:11:06.6 & 0.35 & 89.5 $\pm$ 11.8 & 0.036 $\pm$ 0.003 & 2.59~$\pm$~0.96 & 0$\times$10$^{-5}$ \\
 					& 4 & 22:39:06.5 & 75:11:52.5 & 0.32 & 82.7 $\pm$ 11.0 & 0.068 $\pm$ 0.006 & <~0.78 & <~0$\times$10$^{-5}$ \\
\\
\multirow{2}{*}{ECC276} 	& 1 & 01:38:34.7 & 65:05:48.7 & 0.47 & 16.7 $\pm$ 2.2 & 0.065 $\pm$ 0.006 & <~0.56 & <~0$\times$10$^{-5}$ \\
 					& 2 & 01:38:30.3 & 65:04:52.7 & 0.45 & 18.6 $\pm$ 2.5 & 0.086 $\pm$ 0.009 & <~0.52 & <~0$\times$10$^{-5}$ \\
\\
\multirow{4}{*}{ECC332} 	& 1 & 04:17:23.9 & 55:16:15.7 & 0.20 & 19.8 $\pm$ 2.7 & 0.203 $\pm$ 0.024 & <~6.24 & <~1$\times$10$^{-5}$ \\
 					& 2 & 04:17:04.3 & 55:13:55.0 & 0.08 & 47.2 $\pm$ 6.5 & 0.204 $\pm$ 0.025 & <~0.96 & $\le$~1$\times$10$^{-5}$ \\
 					& 3 & 04:17:02.6 & 55:15:47.0 & 0.15 & 22.1 $\pm$ 3.0 & 0.218 $\pm$ 0.027 & <~3.69 & <~1$\times$10$^{-5}$ \\
 					& 4 & 04:16:57.5 & 55:20:13.2 & 0.16 & 19.2 $\pm$ 2.6 & 0.222 $\pm$ 0.027 & <~4.25 & <~1$\times$10$^{-5}$ \\
\\
\multirow{1}{*}{ECC334} 	& 1 & 04:18:25.6 & 55:12:48.6 & 0.29 & 14.8 $\pm$ 2.0 & 0.196 $\pm$ 0.023 & <~11.35 & <~1$\times$10$^{-5}$ \\
\\
\multirow{3}{*}{ECC335} 	& 1 & 04:18:51.6 & 55:14:44.9 & 0.15 & 24.4 $\pm$ 3.3 & 0.191 $\pm$ 0.022 & <~4.80 & $\le$~1$\times$10$^{-5}$ \\
 					& 2 & 04:19:16.2 & 55:14:02.2 & 0.16 & 28.1 $\pm$ 3.8 & 0.163 $\pm$ 0.018 & <~5.08 & $\le$~1$\times$10$^{-5}$ \\
 					& 3 & 04:19:22.8 & 55:15:54.1 & 0.13 & 33.2 $\pm$ 4.4 & 0.126 $\pm$ 0.013 & <~3.24 & $\le$~1$\times$10$^{-5}$ \\
\\
\multirow{4}{*}{ECC340} 	& 1 & 04:29:27.5 & 54:14:37.5 & 0.07 & 63.7 $\pm$ 8.6 & 0.159 $\pm$ 0.018 & <~0.54 & <~1$\times$10$^{-5}$ \\
 					& 2 & 04:29:48.3 & 54:16:01.8 & 0.08 & 57.8 $\pm$ 8.0 & 0.238 $\pm$ 0.030 & <~0.65 & <~1$\times$10$^{-5}$ \\
 					& 3 & 04:29:41.9 & 54:14:09.6 & 0.08 & 52.7 $\pm$ 7.2 & 0.203 $\pm$ 0.025 & <~0.72 & <~1$\times$10$^{-5}$ \\
 					& 4 & 04:29:51.5 & 54:14:09.7 & 0.09 & 52.6 $\pm$ 7.2 & 0.215 $\pm$ 0.026 & <~0.76 & <~1$\times$10$^{-5}$ \\
\\
\multirow{2}{*}{ECC345} 	& 1 & 04:47:15.7 & 53:01:39.2 & 0.18 & 20.8 $\pm$ 2.9 & 0.171 $\pm$ 0.020 & <~1.70 & $\le$~1$\times$10$^{-5}$ \\
 					& 2 & 04:47:17.2 & 53:04:27.4 & 0.21 & 16.0 $\pm$ 2.2 & 0.183 $\pm$ 0.022 & <~2.40 & <~1$\times$10$^{-5}$ \\
\\
\multirow{2}{*}{ECC346} 	& 1 & 04:48:08.7 & 53:07:23.0 & 0.13 & 78.8 $\pm$ 10.5 & 0.098 $\pm$ 0.010 & <~2.13 & <~1$\times$10$^{-5}$ \\
 					& 2 & 04:48:01.0 & 53:09:15.3 & 0.12 & 83.6 $\pm$ 11.1 & 0.086 $\pm$ 0.008 & <~1.85 & <~1$\times$10$^{-5}$ \\
\\
					
\hline
\label{Table:Clump_Props}
\end{tabular}
\end{center}
\vspace{-0.6cm}
\textbf{Notes}: Size, $\bar{n}$$_{\mathrm{H}}$, $\bar{G}$$_{\mathrm{0}}$, and $S_{1~\mathrm{cm}}^{\mathrm{observed}}$ are taken from~\citet[][]{Tibbs:15a}. 
\end{table*}

%%%%%%% Spinning Dust Emission %%%%%%%%%%%%%%%%%%%%%%%%%%%%%%%%%%%%%%%%%

\section{Spinning Dust Emission}
\label{Sec:Spinning_Dust}

The idea of electric dipole emission from spinning dust grains was first postulated by~\citet{Erickson:57}, and an updated version of this theory was proposed by~\citet{DaL:98} as the source of the observed anomalous microwave emission~\citep[e.g.,][]{Leitch:97}. A competing theory to explain this anomalous microwave emission is magnetic dipole emission~\citep{DaL:99, Draine:13}, and a recent analysis by~\citet{Hensley:15} found a lack of correlation between the anomalous microwave emission and the presence of very small grains, casting doubt on the spinning dust mechanism and highlighting the possible importance of magnetic dipole emission. Although the situation is not completely clear, polarisation observations~\citep[e.g.,][]{Dickinson:11} appear to favour the spinning dust hypothesis, and therefore throughout this analysis we ignore any possible contribution from magnetic dipole emission.

Since~\citet{DaL:98} published their model, the spinning dust model has been updated and refined, including a quantum mechanical treatment~\citep{Ysard:10b}, incorporating the impact of ion collisions~\citep{Hoang:10} and accounting for irregular dust grain shapes~\citep{Hoang:11}. In addition to these updates,~\citet{Ali-Haimoud:09} used the Fokker-Planck equation to explicitly determine the angular velocity distribution rather than assuming a Maxwellian distribution and made the model public in the form of \textsc{spdust}\footnote{http://www.sns.ias.edu/$\sim$yacine/spdust/spdust.html}, which~\citet{Silsbee:11} later updated to account for dust grains rotating about a non-principal axis. In this analysis we use this updated version of \textsc{spdust}. 

Although there are slight variations between the different spinning dust models, all of the models agree on the importance of the dust grain size distribution.

\begin{figure*}
\begin{center}
\includegraphics[angle=0,scale=0.60]{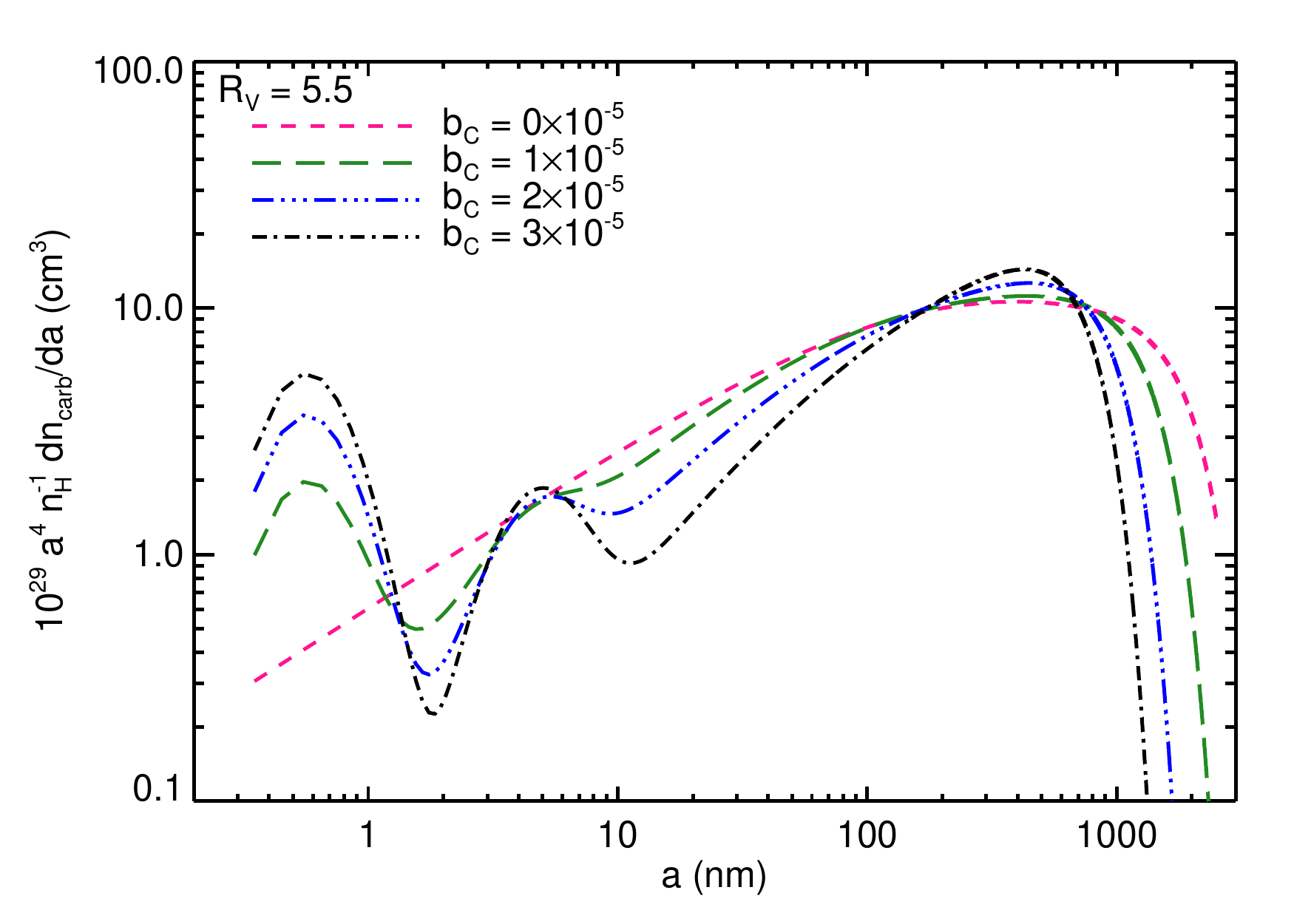}
\end{center}
\vspace{-0.3cm}
\caption{Grain size distribution for carbonaceous dust grains incorporated within \textsc{spdust}. The distribution is plotted for four different values of $b_{\mathrm{C}}$, the very small dust grain abundance.}
\label{Fig:Size_Dist}
\end{figure*}

%%%%%%% Size Distribution %%%%%%%%%%%%%%%%%%%%%%%%%%%%%%%%%%%%%%%%%%%%

\subsection{Dust Grain Size Distribution}
\label{Subsec:Size_Dist}

\textsc{spdust} incorporates the~\citet{Weingartner:01} size distribution for carbonaceous dust grains with the functional form: 

\begin{equation}
\begin{split}
\frac{1}{n_{\mathrm{H}}} \frac{dn_{\mathrm{carb}}}{da} = D(a) + \frac{C_{\mathrm{g}}}{a} \left( \frac{a}{a_{\mathrm{t,g}}} \right)^{\alpha_{\mathrm{g}}} F(a; \beta_{\mathrm{g}}, a_{\mathrm{t,g}}) \\
\times
\begin{cases}
1, 												& 	0.35~\mathrm{nm} < a < a_{\mathrm{t,g}} \\
\mathrm{exp}~\{ -[(a-a_{\mathrm{t,g}})/a_{\mathrm{c,g}}]^{3} \}, 	& 	a > a_{\mathrm{t,g}}
\end{cases}
,
\end{split}
\label{equ:dnda}
\end{equation}

\noindent
where

\begin{equation}
F(a; \beta_{\mathrm{g}}, a_{\mathrm{t,g}}) \equiv 
\begin{cases}
1 + \beta_{\mathrm{g}} a/a_{\mathrm{t,g}}, 			&	\beta_{\mathrm{g}} \ge 0 \\
(1 - \beta_{\mathrm{g}} a/a_{\mathrm{t,g}})^{-1},		&	\beta_{\mathrm{g}} < 0
\end{cases}
,
\label{equ:F}
\end{equation}

\noindent
where the function $D(a)$ in Equation~\eqref{equ:dnda} accounts for the size distribution of the population of very small grains. Observations of thermal dust emission at wavelengths between~$\sim$3~--~60~$\mu$m invoked a population of very small grains that was larger than that which could be obtained by extrapolating the~\citet{Mathis:77} size distribution~(modelled as a power-law of the form $dn/da \propto a^{-3.5}$) to very small sizes~\citep[e.g.,][]{Desert:90}. Therefore, a population of very small grains was required that exhibited a size range so that the absorption of a single photon could raise the temperature sufficiently, causing the grains to emit at wavelengths of~$\sim$3~--~60~$\mu$m. After comparing observations of diffuse Galactic emission with detailed models,~\citet{Li:01} concluded that the IR emission can be well reproduced if the population of very small dust grains is modelled as the sum of two log-normal size distributions:

\begin{equation}
D(a) = \sum\limits_{i = 1}^{2} \frac{B_{i}}{a} \mathrm{exp} \left\{ -\frac{1}{2} \left[ \frac{\mathrm{ln}(a/a_{\mathrm{0,}i})}{\sigma} \right]^{2} \right\}, a > 0.35~\mathrm{nm} ,
\label{equ:Da}
\end{equation}

\noindent
where

\begin{equation}
\begin{split}
B_{i} = \frac{3}{(2\pi)^{3/2}} \frac{\mathrm{exp}~(-4.5 \sigma^{2})}{\rho_{g} a_{\mathrm{0,}i}^{3}\sigma} \\
\times \frac{b_{\mathrm{C,}i}m_{\mathrm{C}}}{1 + \mathrm{erf}~[ 3 \sigma/\sqrt 2 + \mathrm{ln}(a_{\mathrm{0,}i}/0.35~\mathrm{nm})/\sigma \sqrt2 ]} ,
\end{split}
\label{equ:B}
\end{equation}

\noindent
and

\begin{equation}
b_{\mathrm{C}} = \sum\limits_{i = 1}^{2} b_{\mathrm{C},i} ,
\label{equ:bc}
\end{equation}

\noindent
where $m_{\mathrm{C}}$ is the mass of a C atom, $\rho_{g}$ = 2.24~g~cm$^{-3}$ is the mass density of graphite, $a_{\mathrm{0,}i}$ denotes the location of the log-normal peak, $\sigma$ determines the width of the log-normal peak, $b_{\mathrm{C,1}}$ = 0.75$b_{\mathrm{C}}$, $b_{\mathrm{C,2}}$ = 0.25$b_{\mathrm{C}}$, $b_{\mathrm{C}}$ is the total number of C atoms per H nucleus in the log-normal populations, $a_{\mathrm{0,1}}$ = 0.35~nm, $a_{\mathrm{0,2}}$ = 3~nm, and $\sigma$ = 0.4. The log-normal distribution with $a_{\mathrm{0,1}}$ = 0.35~nm is required to reproduce the observed emission at wavelengths of~$\sim$3~--~25~$\mu$m, while the $a_{\mathrm{0,2}}$ = 3~nm component is required to reproduce the observed emission at wavelengths of~$\sim$60~$\mu$m. 

The values of the adjustable parameters for the size distribution of carbonaceous dust grains~($C_{\mathrm{g}}$, $a_{\mathrm{t,g}}$, $\alpha_{\mathrm{g}}$, $a_{\mathrm{c,g}}$, $\beta_{\mathrm{g}}$) are given in Table 1 of~\citet{Weingartner:01} for three different values of $R_{\mathrm{V}}$~=~3.1, 4.0, and 5.5 for a range of values of $b_{\mathrm{C}}$. Although \citet{Weingartner:01} were able to use this size distribution to fit observations for a range of $R_{\mathrm{V}}$ values, they could not accurately constrain $b_{\mathrm{C}}$, and in this analysis we show that by using spinning dust emission, we can provide a constraint on $b_{\mathrm{C}}$.

For the dense cores in which we are focusing on in this analysis, a value of $R_{\mathrm{V}}$ = 5.5 is most appropriate, as has been observed in other dense molecular environments~\citep[][]{Kandori:03, Foster:13}. Therefore, we used only those dust grain size distribution parameters estimated for $R_{\mathrm{V}}$ = 5.5. In Figure~\ref{Fig:Size_Dist} we plot the full size distribution described by Equations~\eqref{equ:dnda},~\eqref{equ:F},~\eqref{equ:Da},~\eqref{equ:B}, and~\eqref{equ:bc} for four values of the very small dust grain abundance: $b_{\mathrm{C}}$ = 0$\times$10$^{-5}$; $b_{\mathrm{C}}$ = 1$\times$10$^{-5}$; $b_{\mathrm{C}}$ = 2$\times$10$^{-5}$; $b_{\mathrm{C}}$ = 3$\times$10$^{-5}$. This plot illustrates how the size distribution, in particular the two log-normal components representing the very small grains, depends on $b_{\mathrm{C}}$. 

Within \textsc{spdust}, grains with $a$~$\le$~0.6~nm~(equivalent to $\lesssim$~100 C atoms) are assumed to be planar and disc-like, while larger grains are treated as spherical.

\begin{figure*}
\begin{center}
\includegraphics[angle=0,scale=0.450]{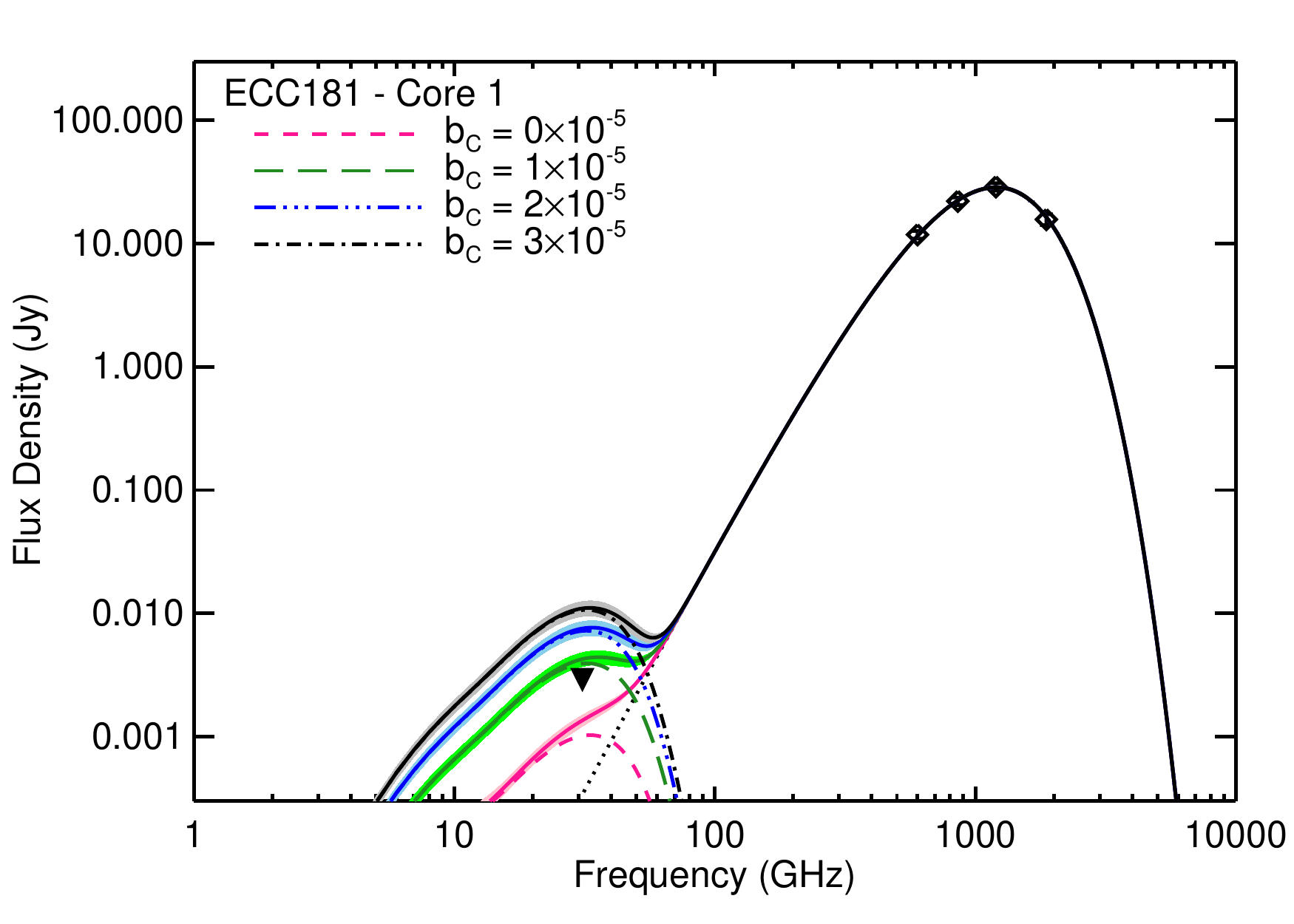}
\includegraphics[angle=0,scale=0.450]{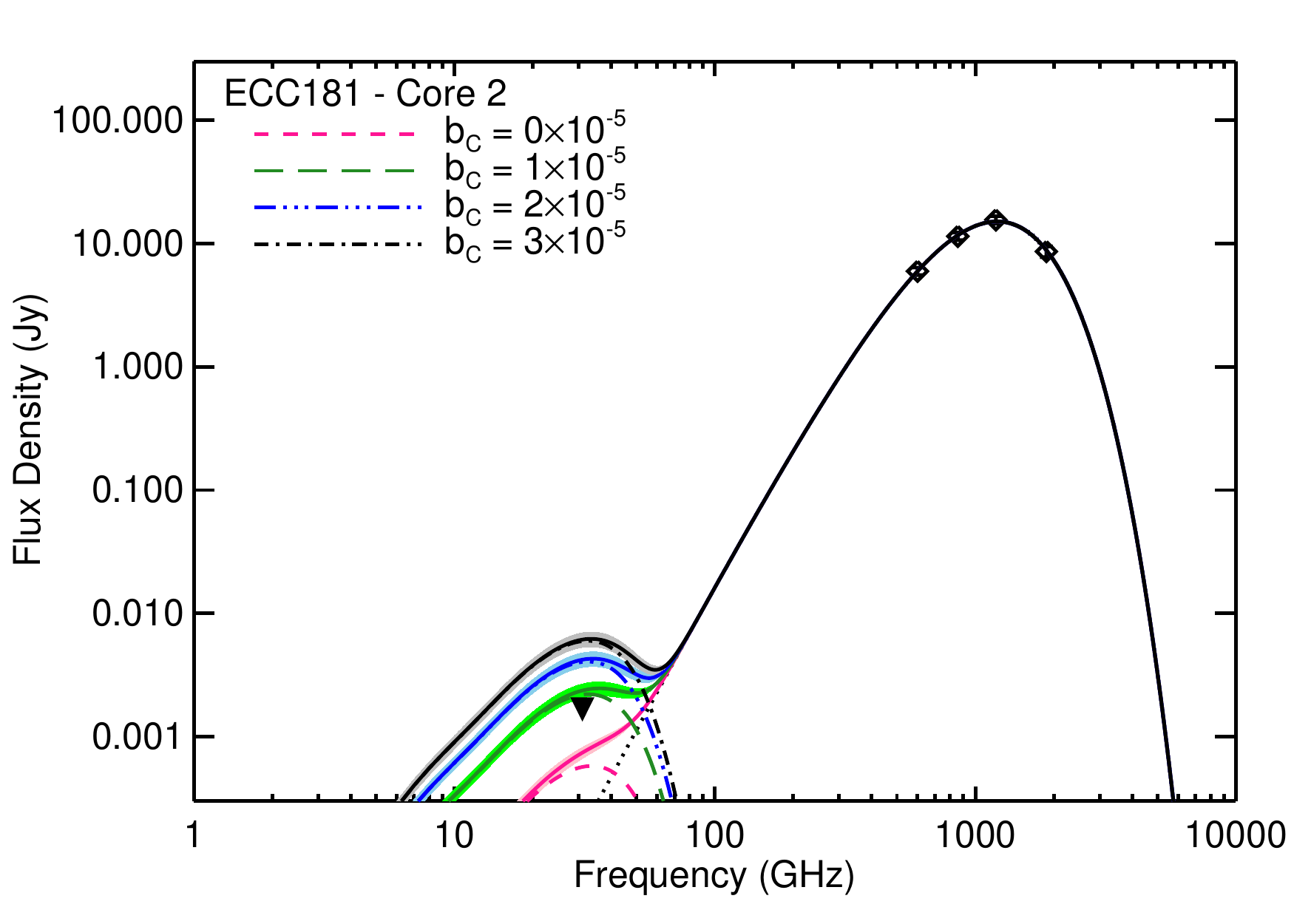} \\
\vspace{-0.35cm}
\includegraphics[angle=0,scale=0.450]{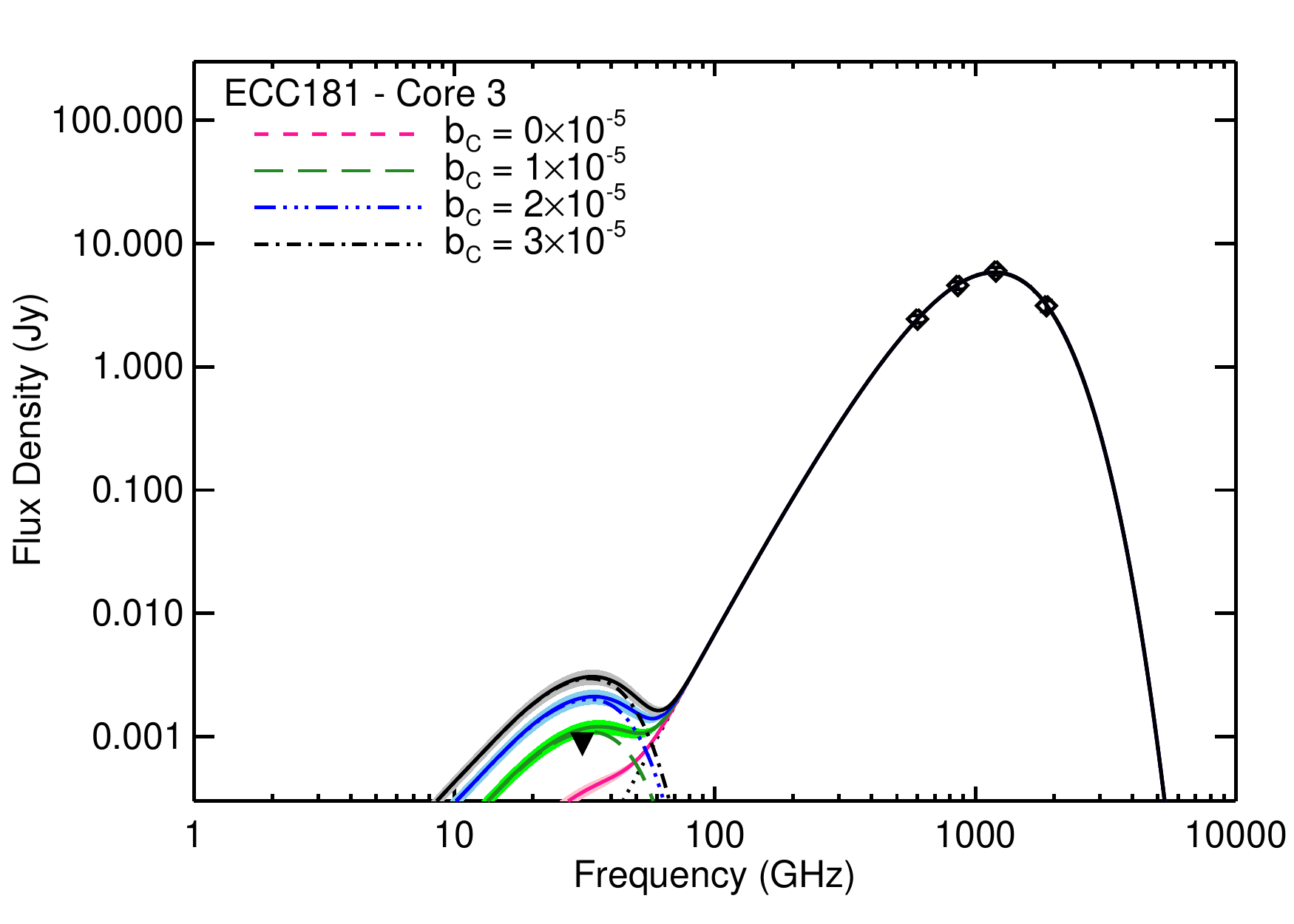} 
\includegraphics[angle=0,scale=0.450]{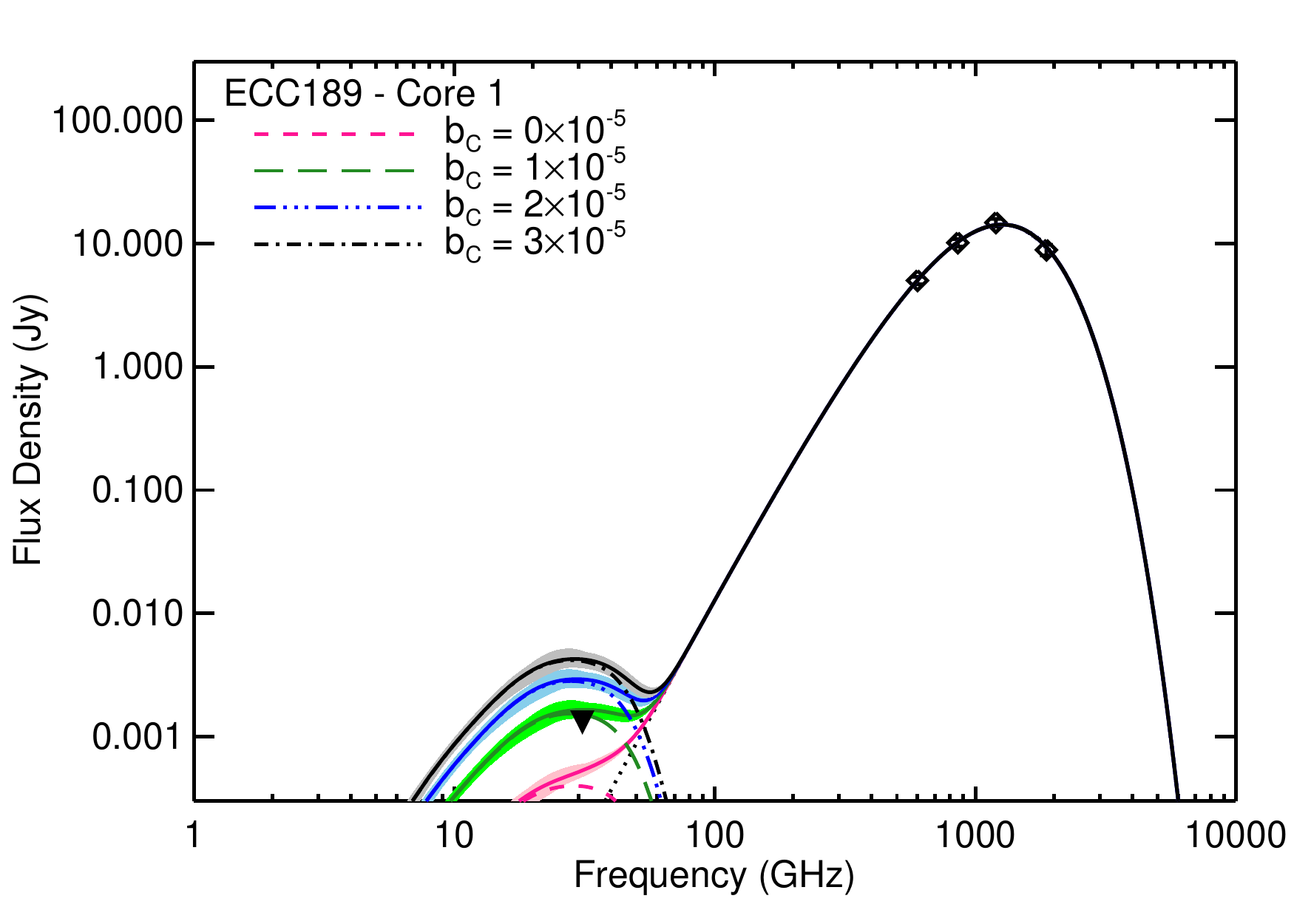} \\
\vspace{-0.35cm}
\includegraphics[angle=0,scale=0.450]{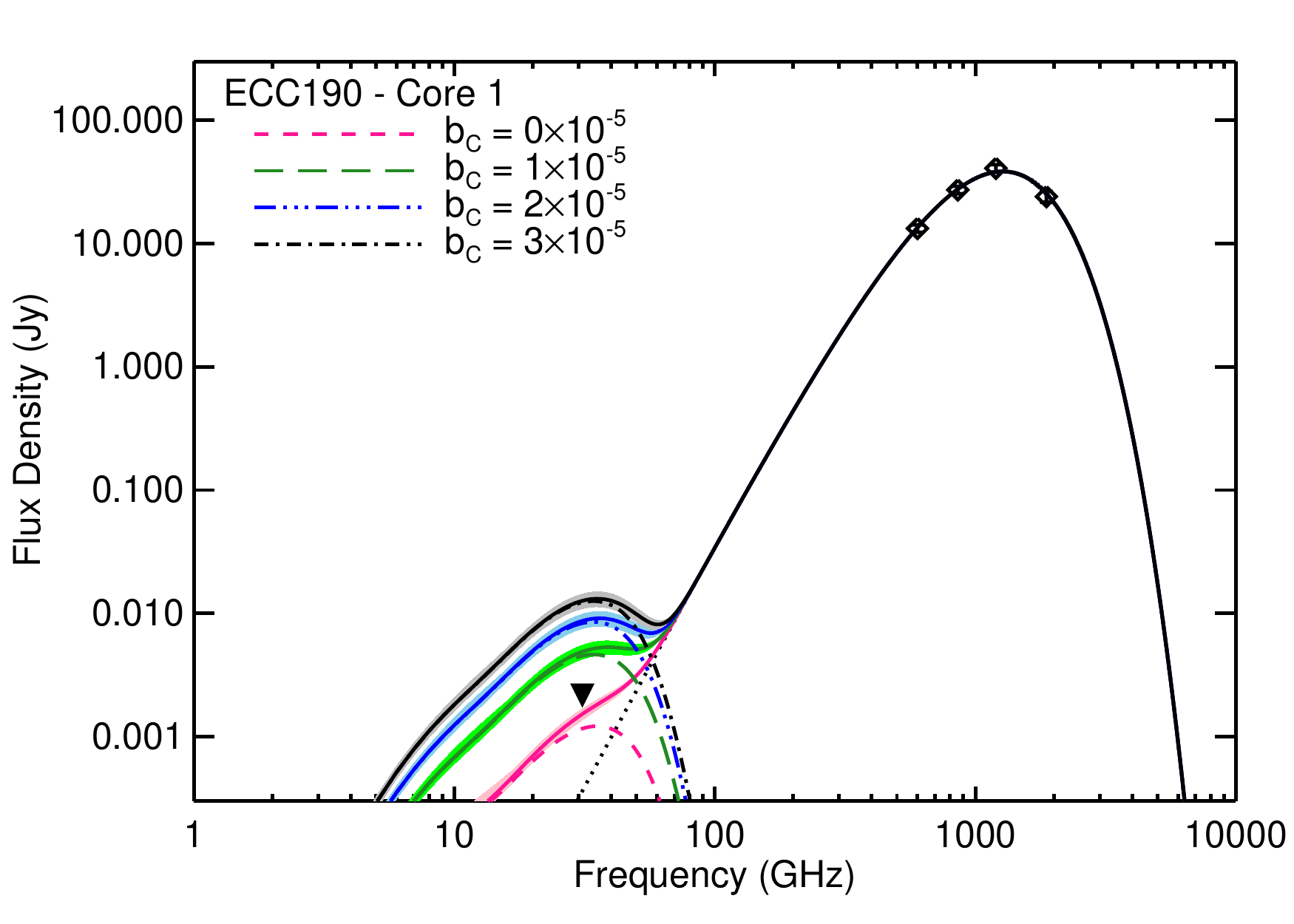} 
\includegraphics[angle=0,scale=0.450]{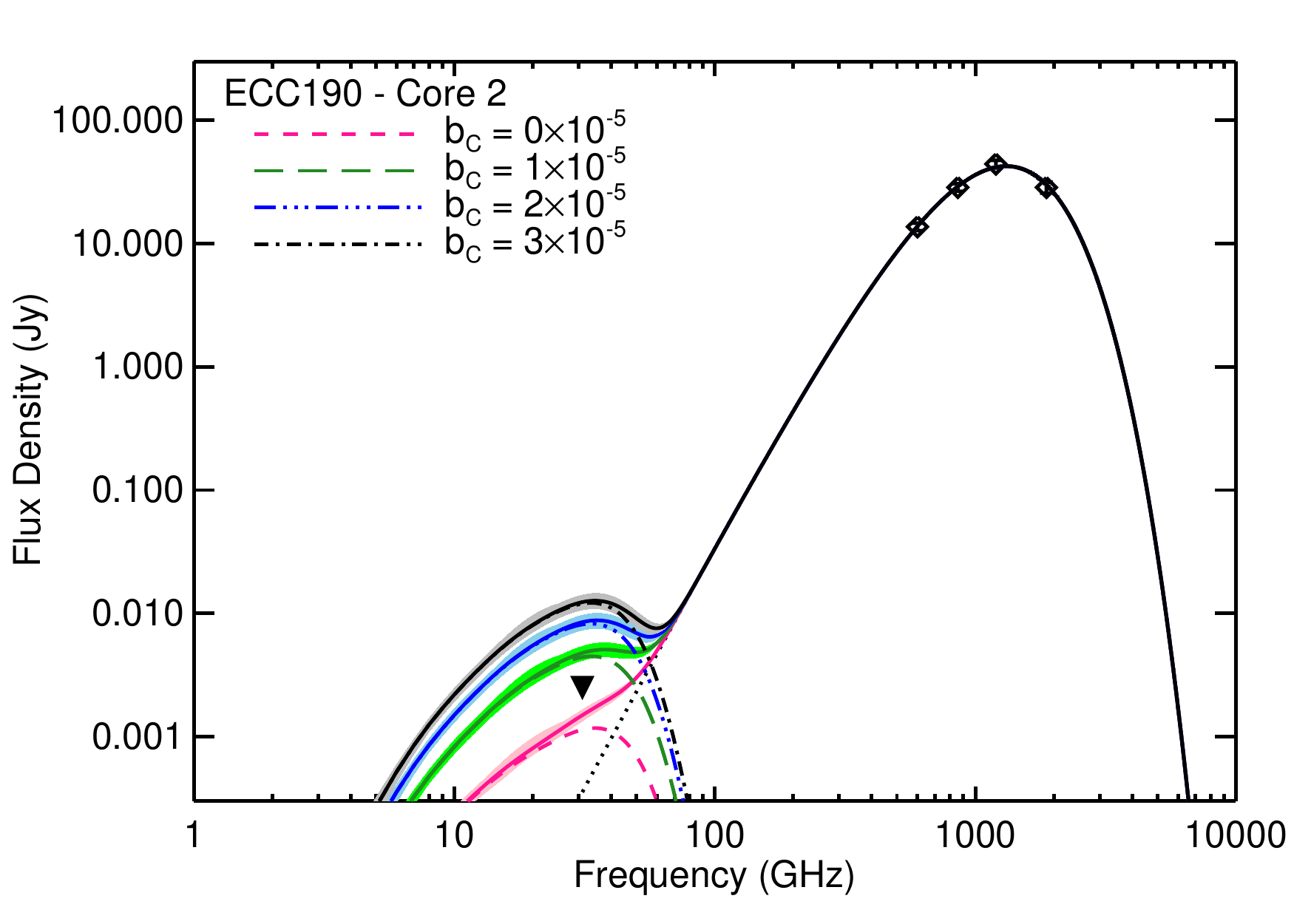} \\
\vspace{-0.35cm}
\includegraphics[angle=0,scale=0.450]{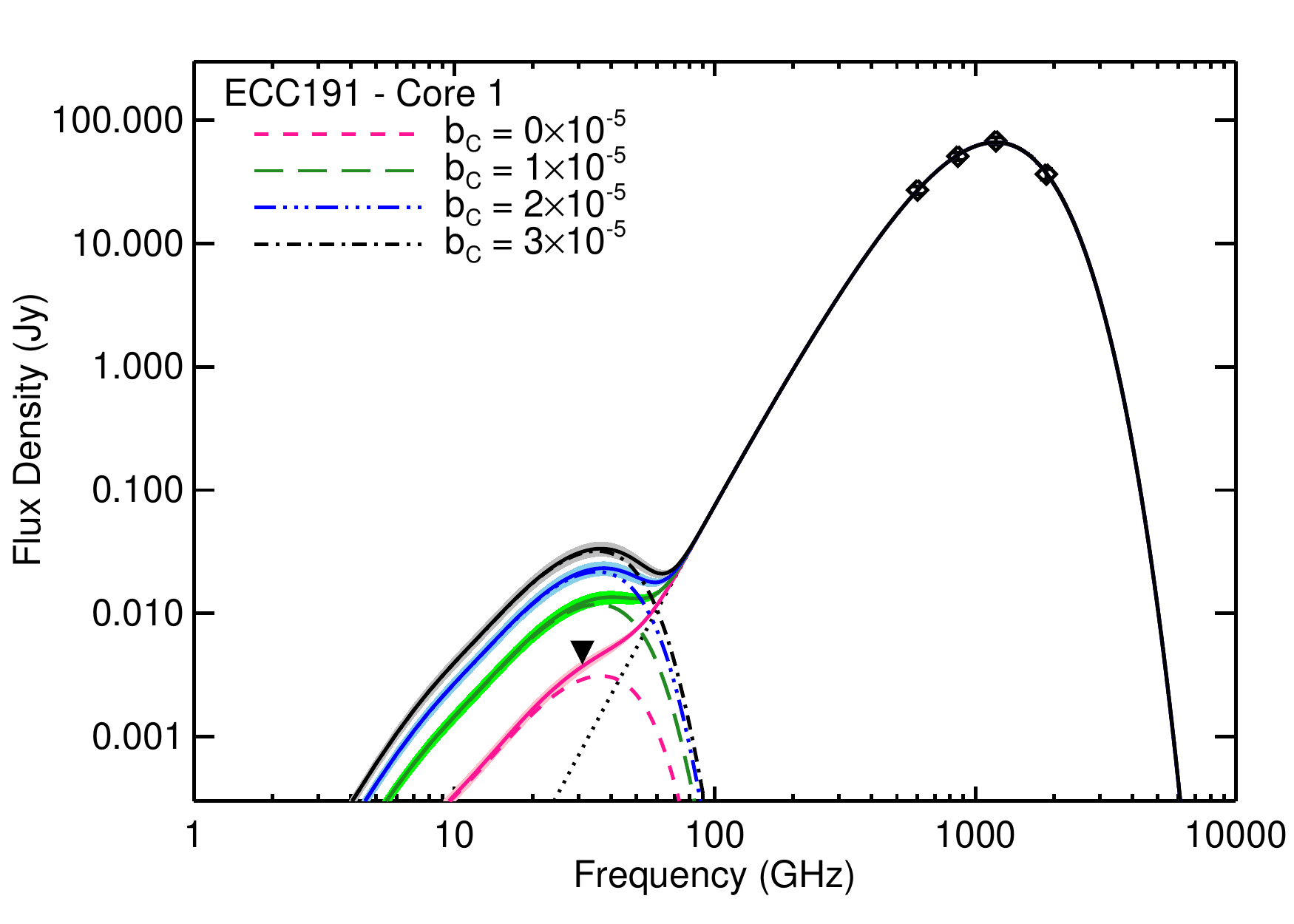} 
\includegraphics[angle=0,scale=0.450]{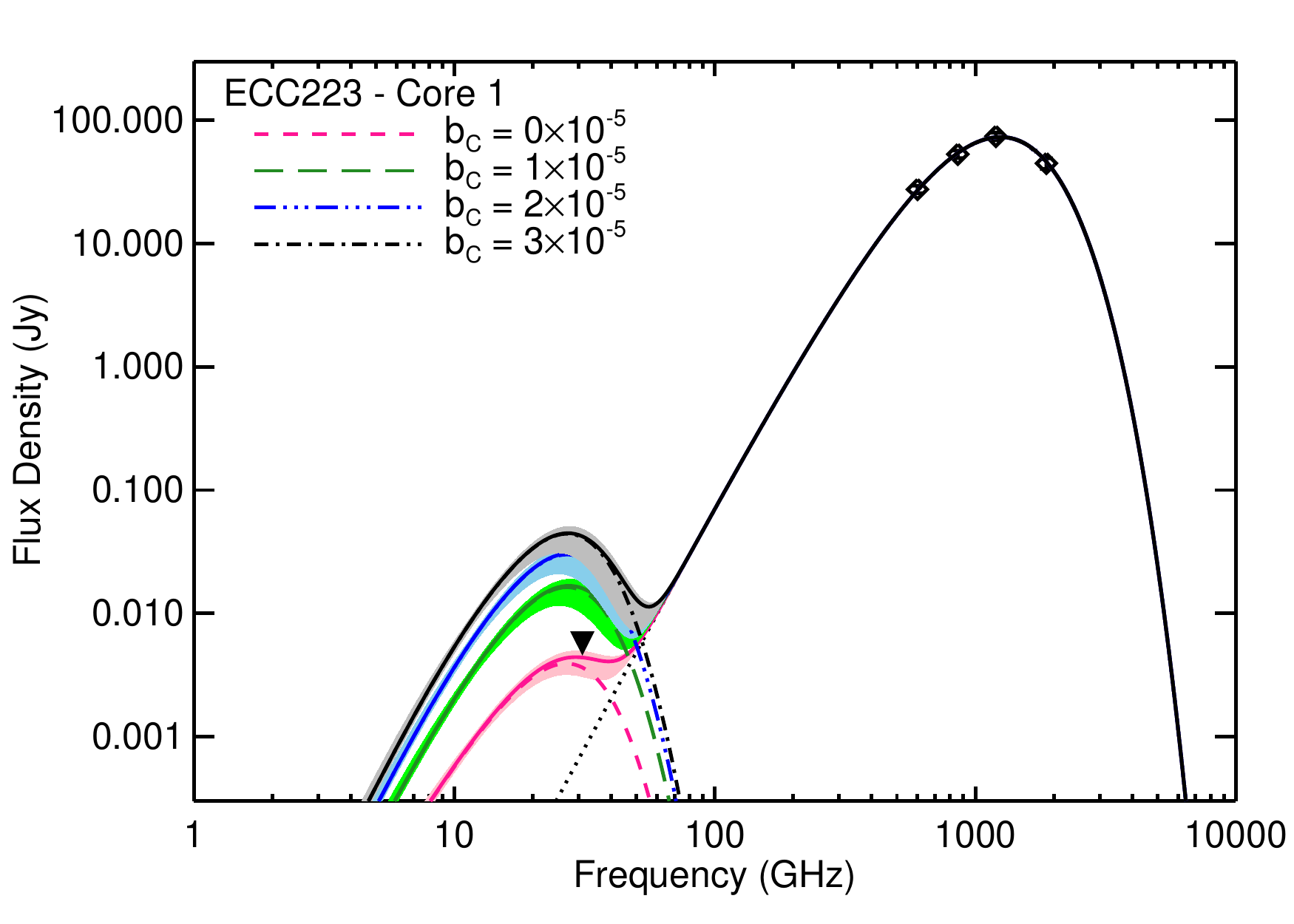} \\
\end{center}
\vspace{-0.30cm}
\caption{Spectra for all of the cores. The integrated flux density at 160, 250, 350, and 500~$\mu$m~(\textit{diamonds}) in each core is fitted with a modified black body~(\textit{dotted line}) with $\beta$ fixed at 2. The predicted spinning dust emission, including the associated uncertainty, is plotted for four different $b_{\mathrm{C}}$ values. The solid curves represent the total emission (spinning dust $+$ thermal dust). Also plotted is the measured flux density at 1~cm~(for sources in which we do not detect any cm emission, we have plotted the 5$\sigma$ upper limit). These plots clearly illustrate that there is a deficit of very small grains in these cores, with values constrained to be $b_{\mathrm{C}}$~$\le$~1$\times$10$^{-5}$.}
\label{Fig:SED_sub_clumps}
\end{figure*}

\begin{figure*}
\ContinuedFloat
\begin{center}
\includegraphics[angle=0,scale=0.450]{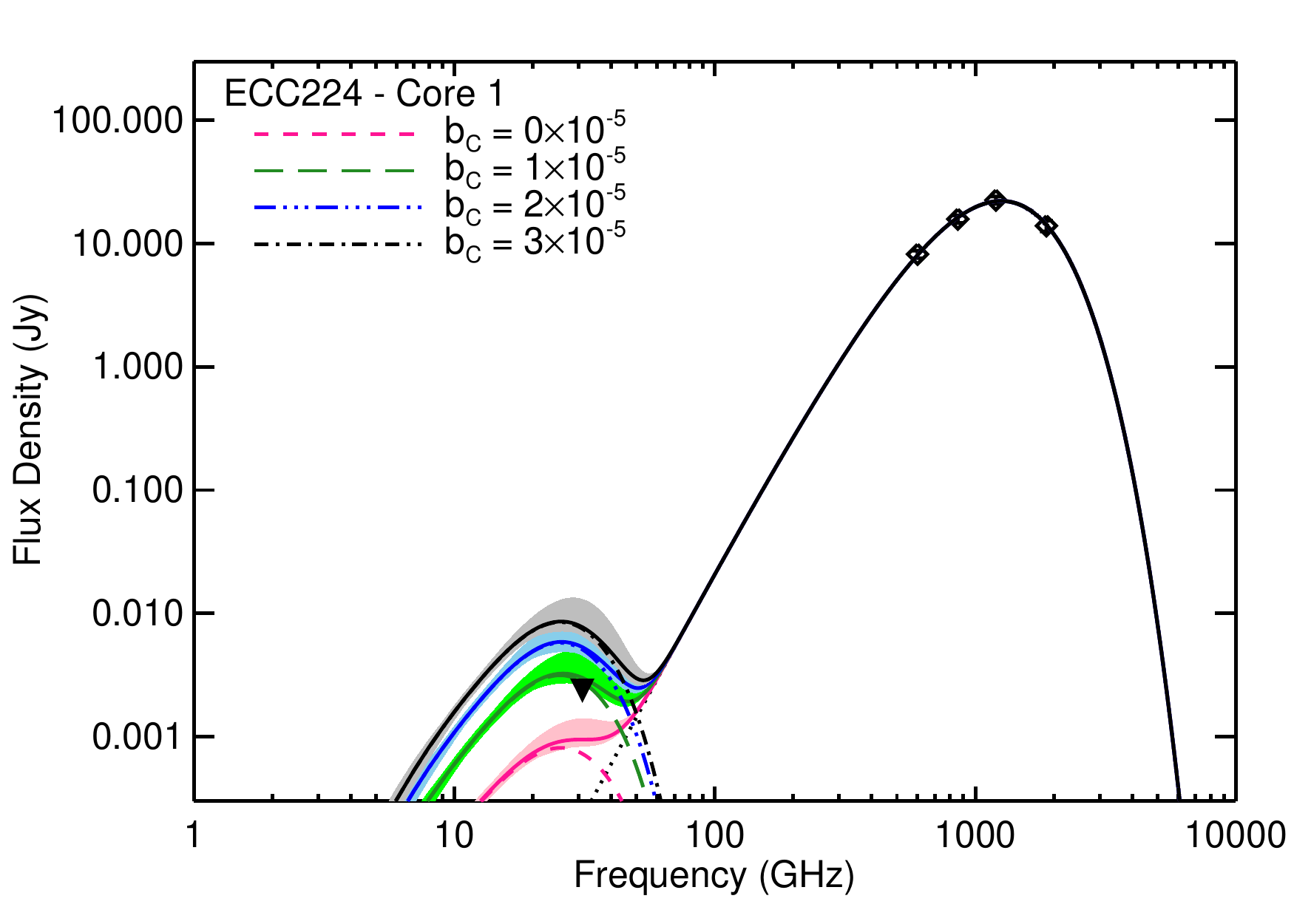}
\includegraphics[angle=0,scale=0.450]{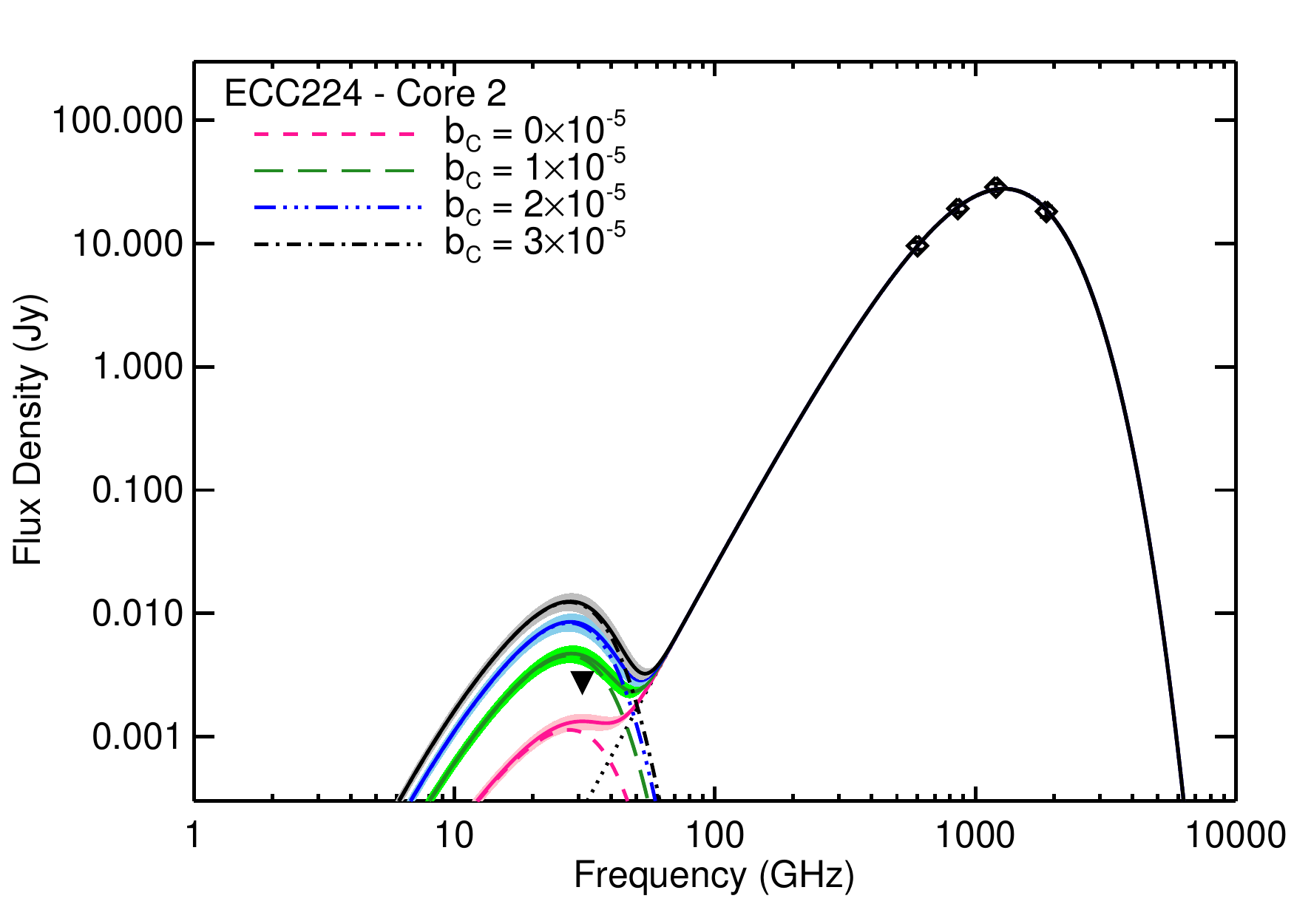} \\
\vspace{-0.35cm}
\includegraphics[angle=0,scale=0.450]{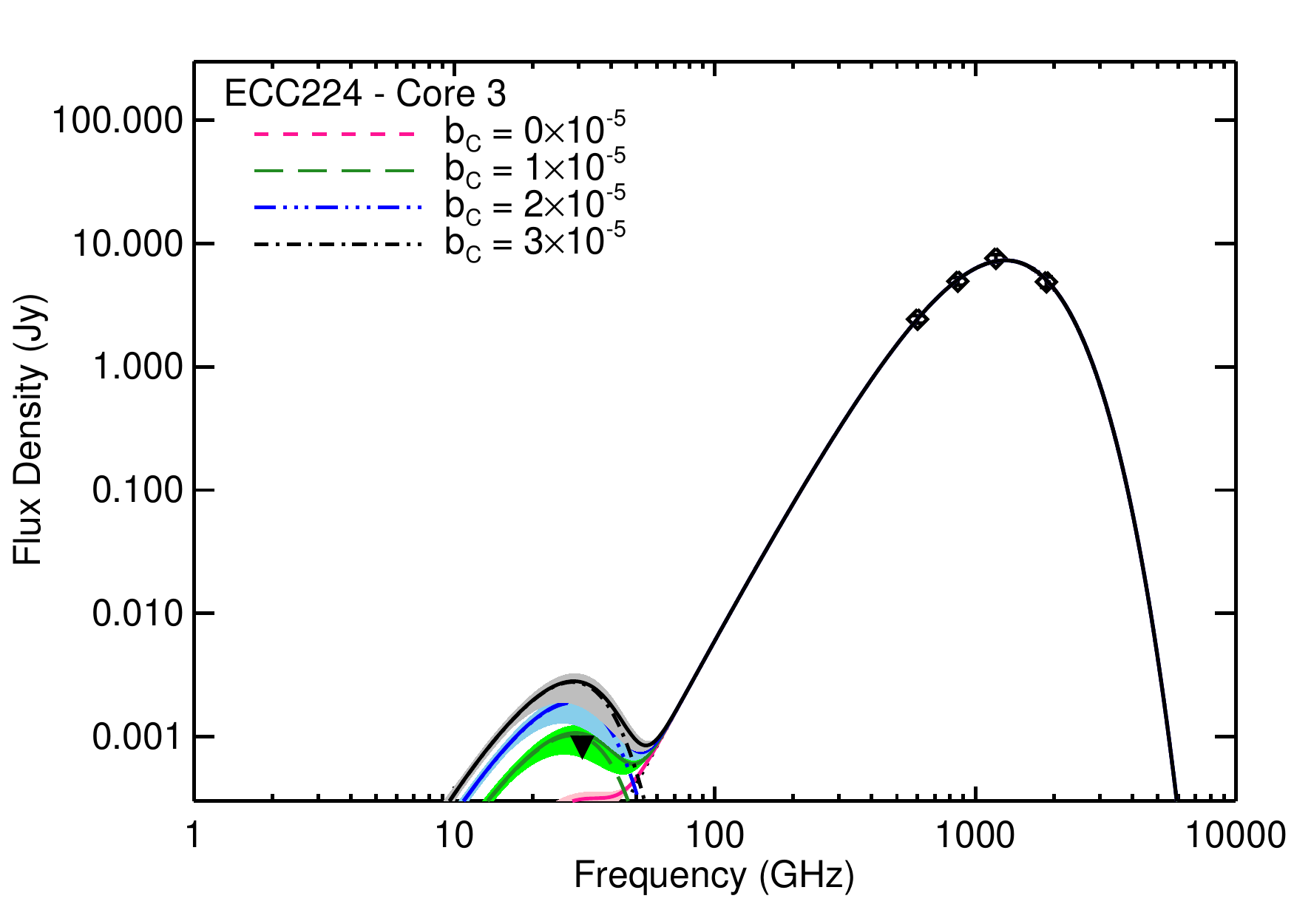}
\includegraphics[angle=0,scale=0.450]{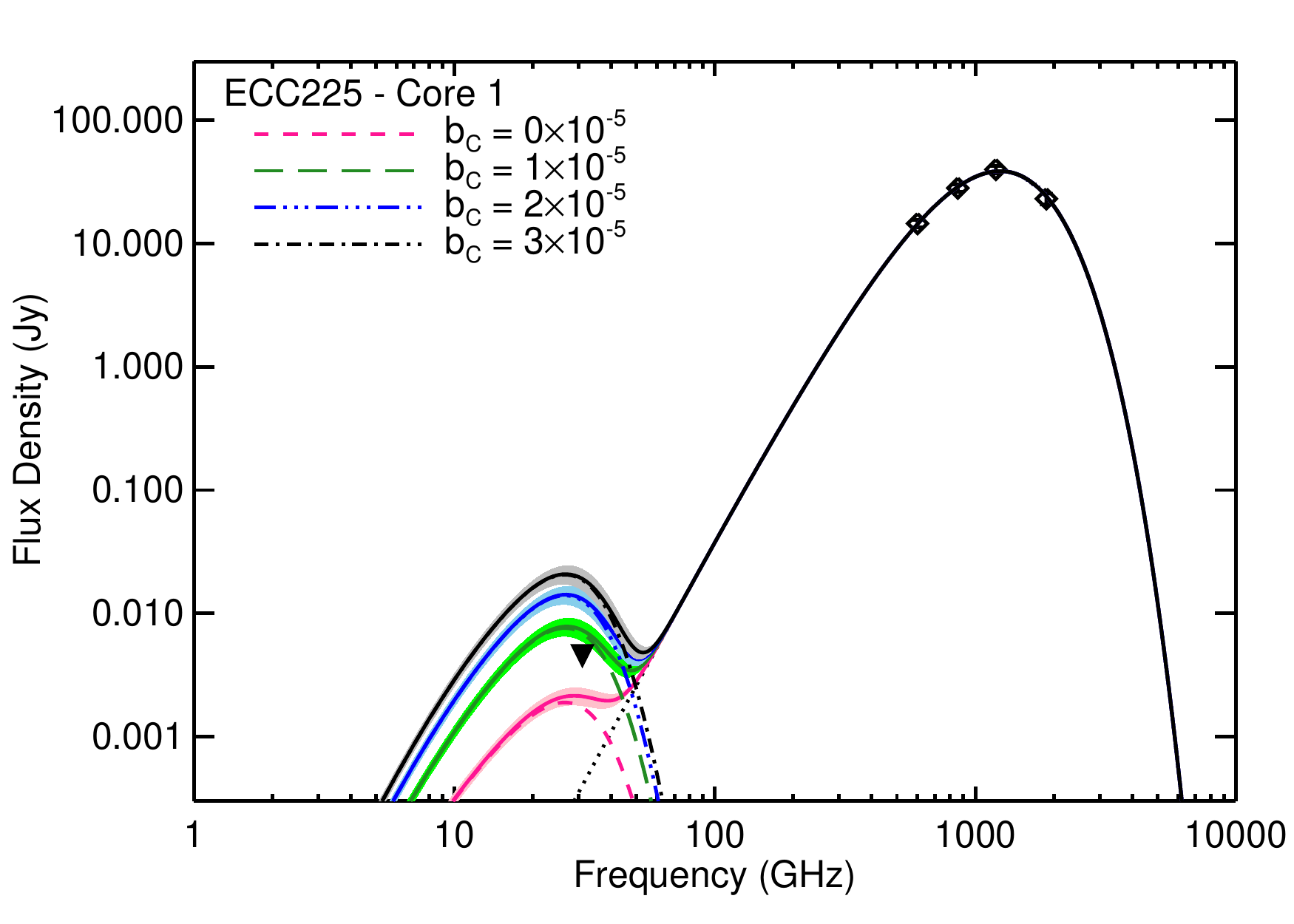} \\
\vspace{-0.35cm}
\includegraphics[angle=0,scale=0.450]{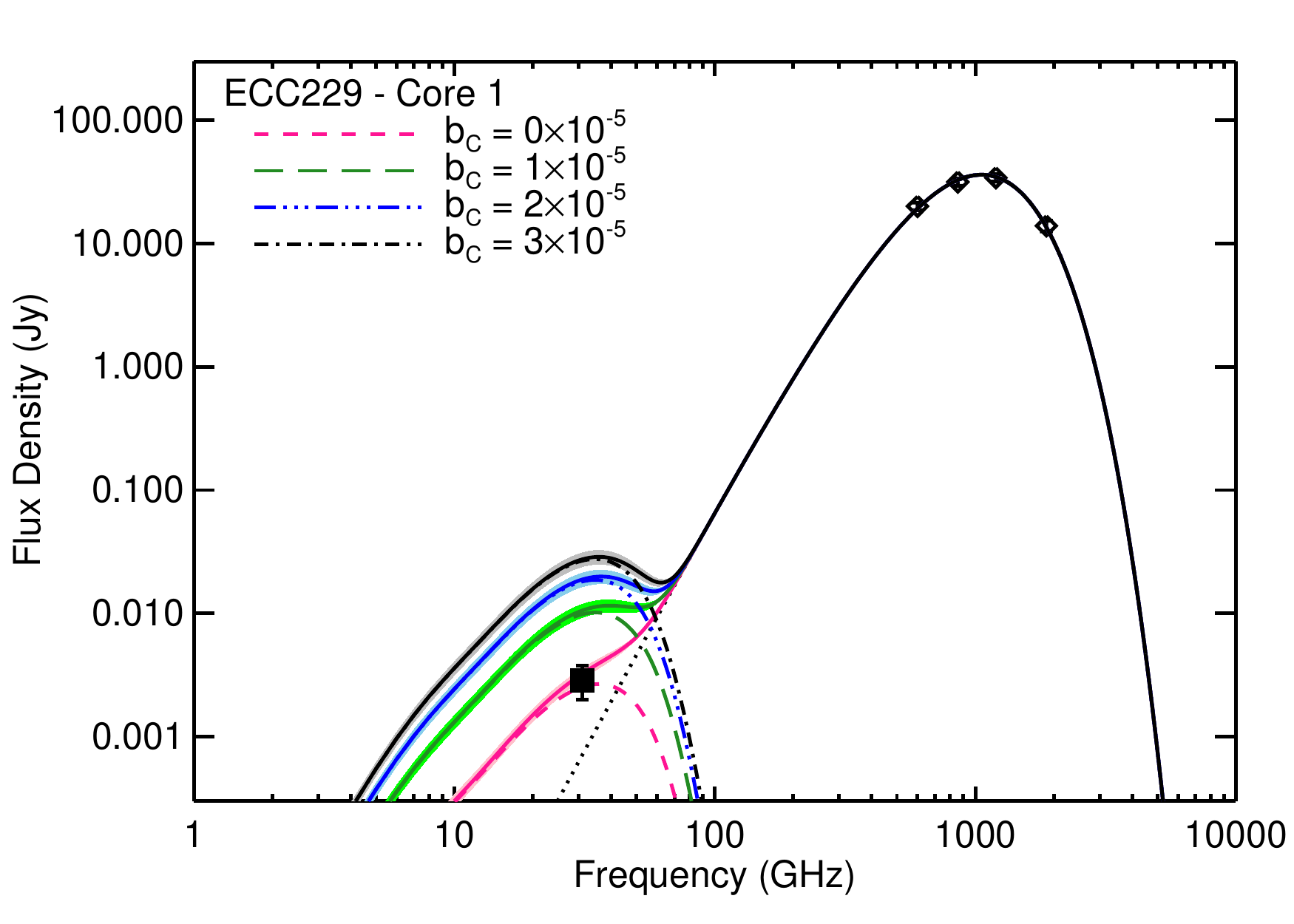}
\includegraphics[angle=0,scale=0.450]{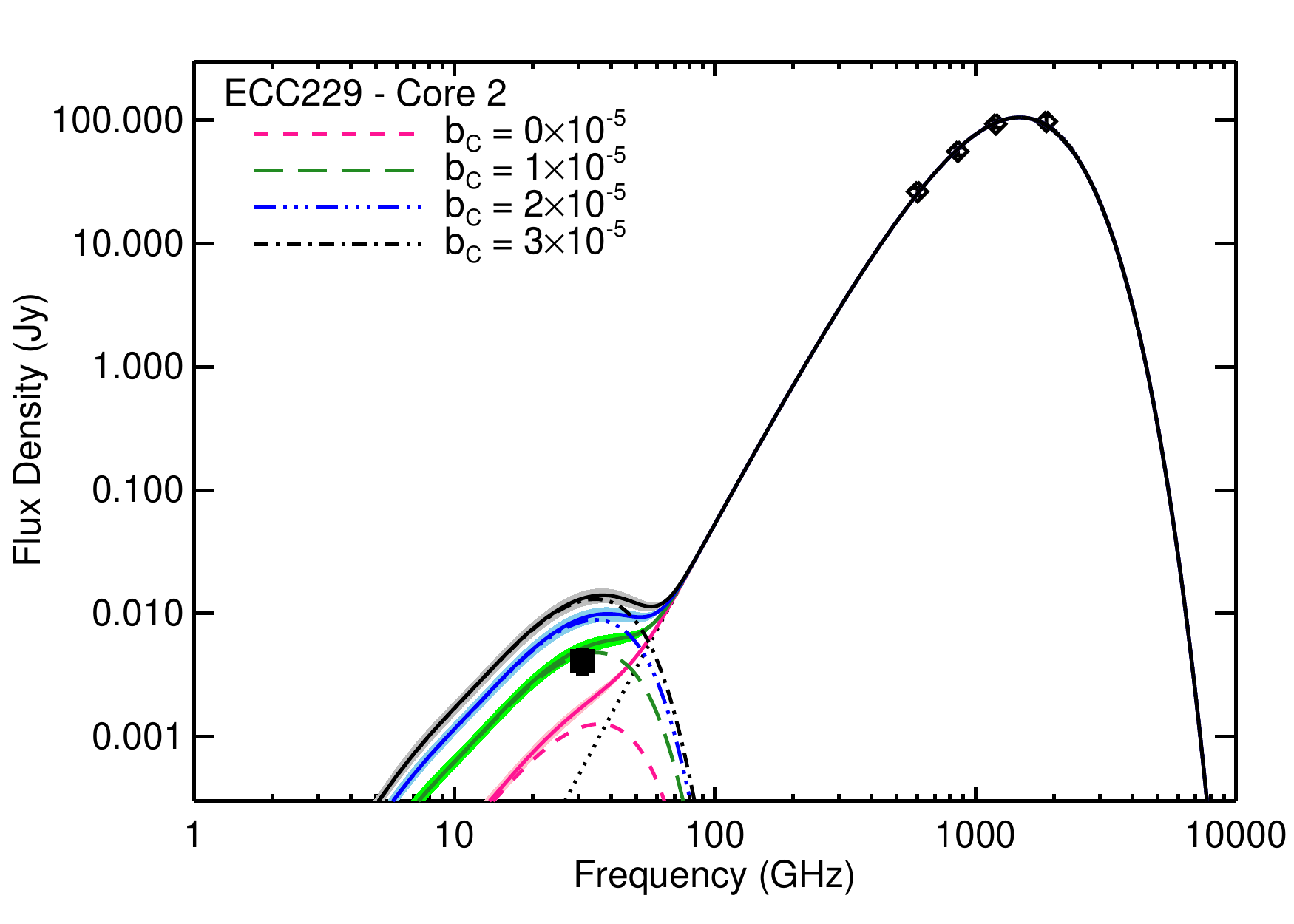} \\
\vspace{-0.35cm}
\includegraphics[angle=0,scale=0.450]{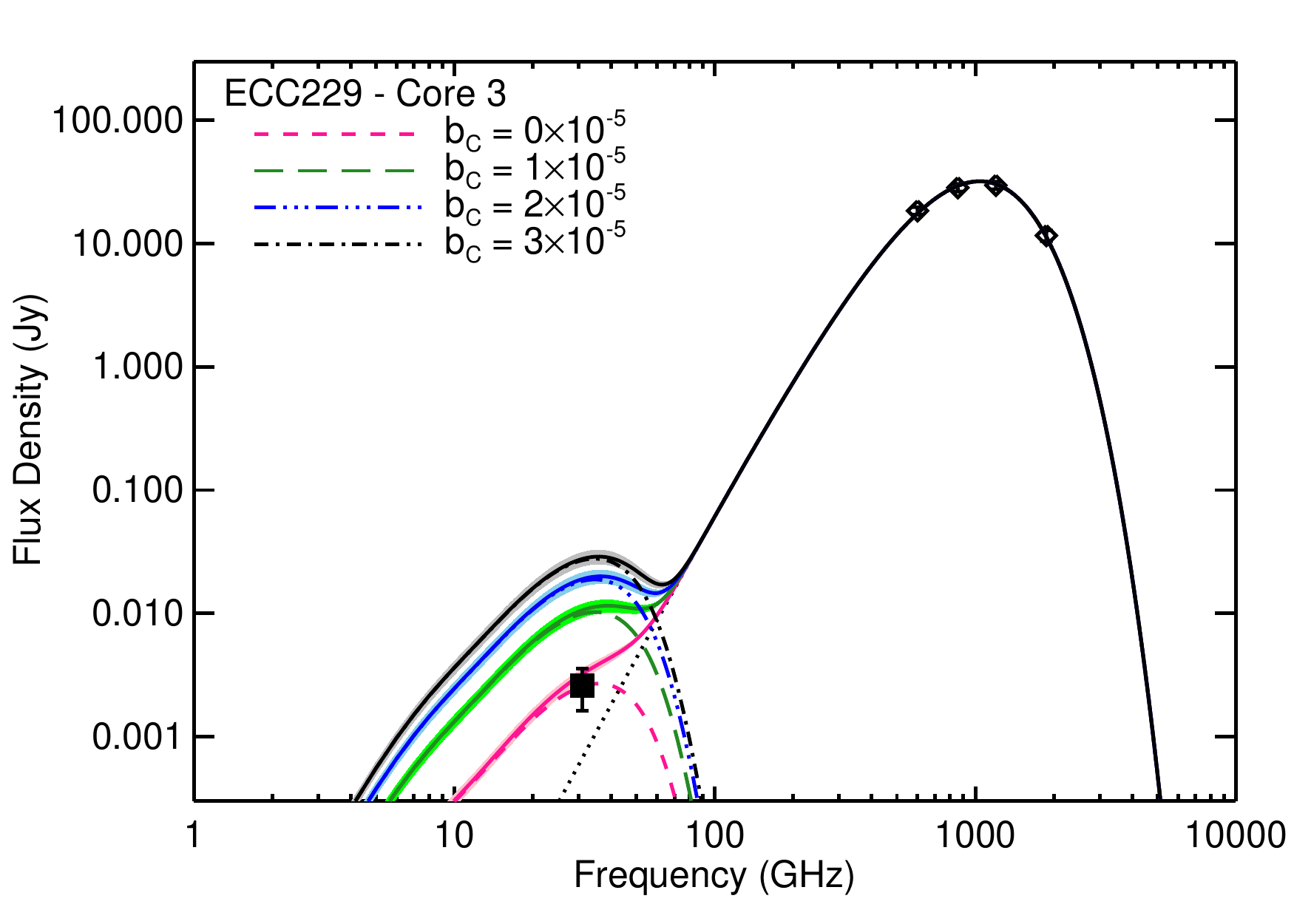} 
\includegraphics[angle=0,scale=0.450]{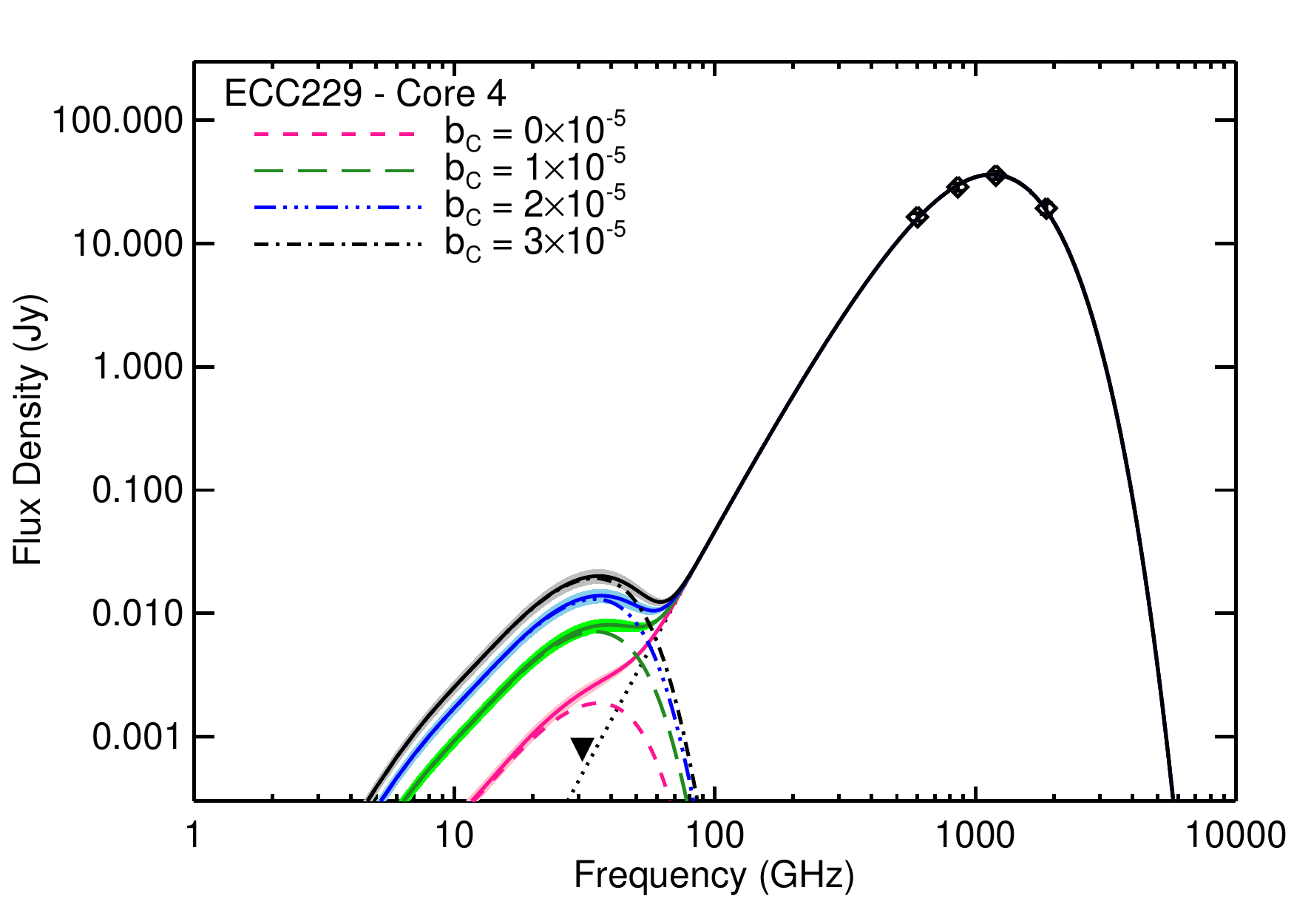}
\end{center}
\vspace{-0.30cm}
\caption{Continued}
\label{Fig:SED_sub_clumps}
\end{figure*}

\begin{figure*}
\ContinuedFloat
\begin{center}
\includegraphics[angle=0,scale=0.450]{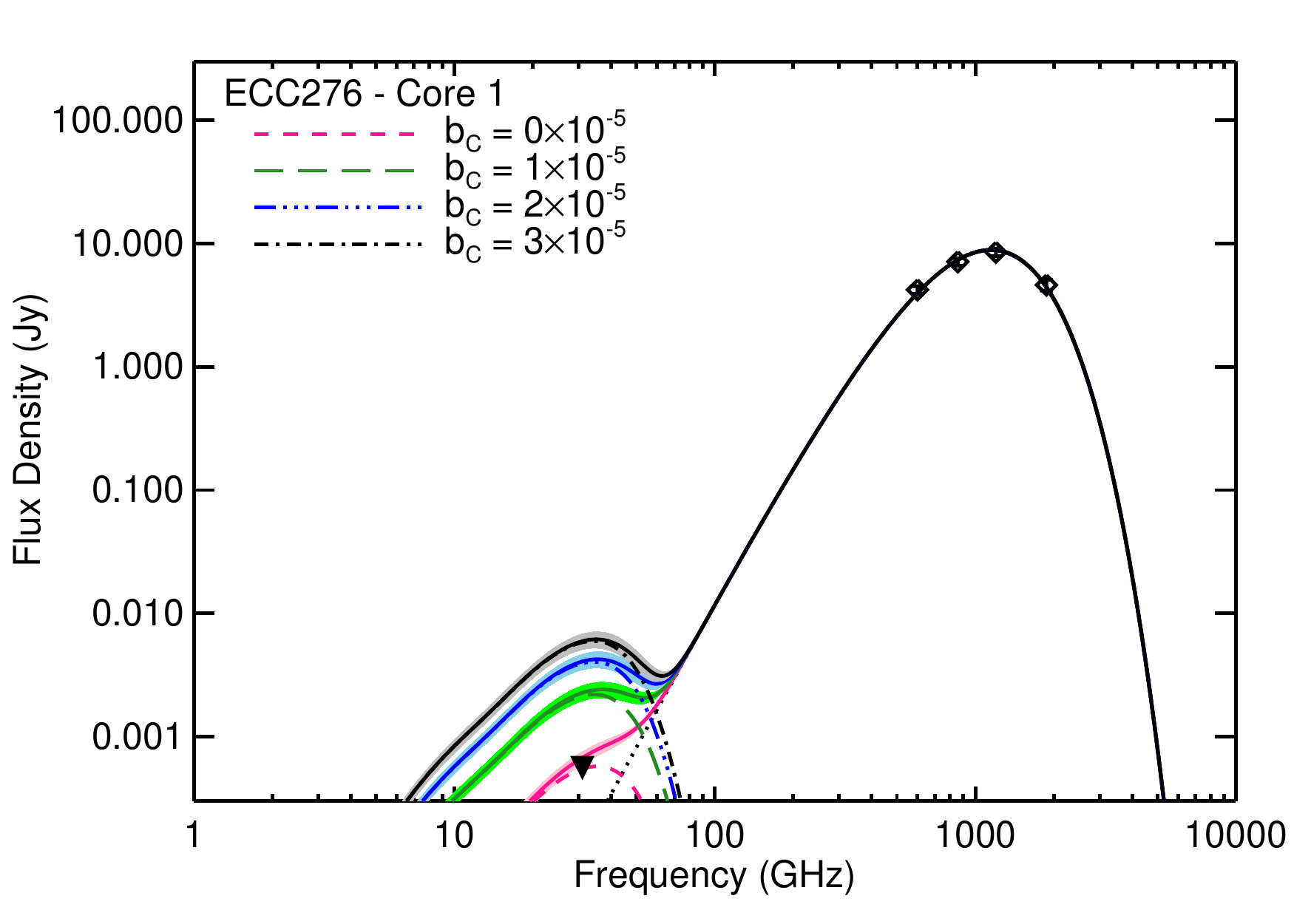}
\includegraphics[angle=0,scale=0.450]{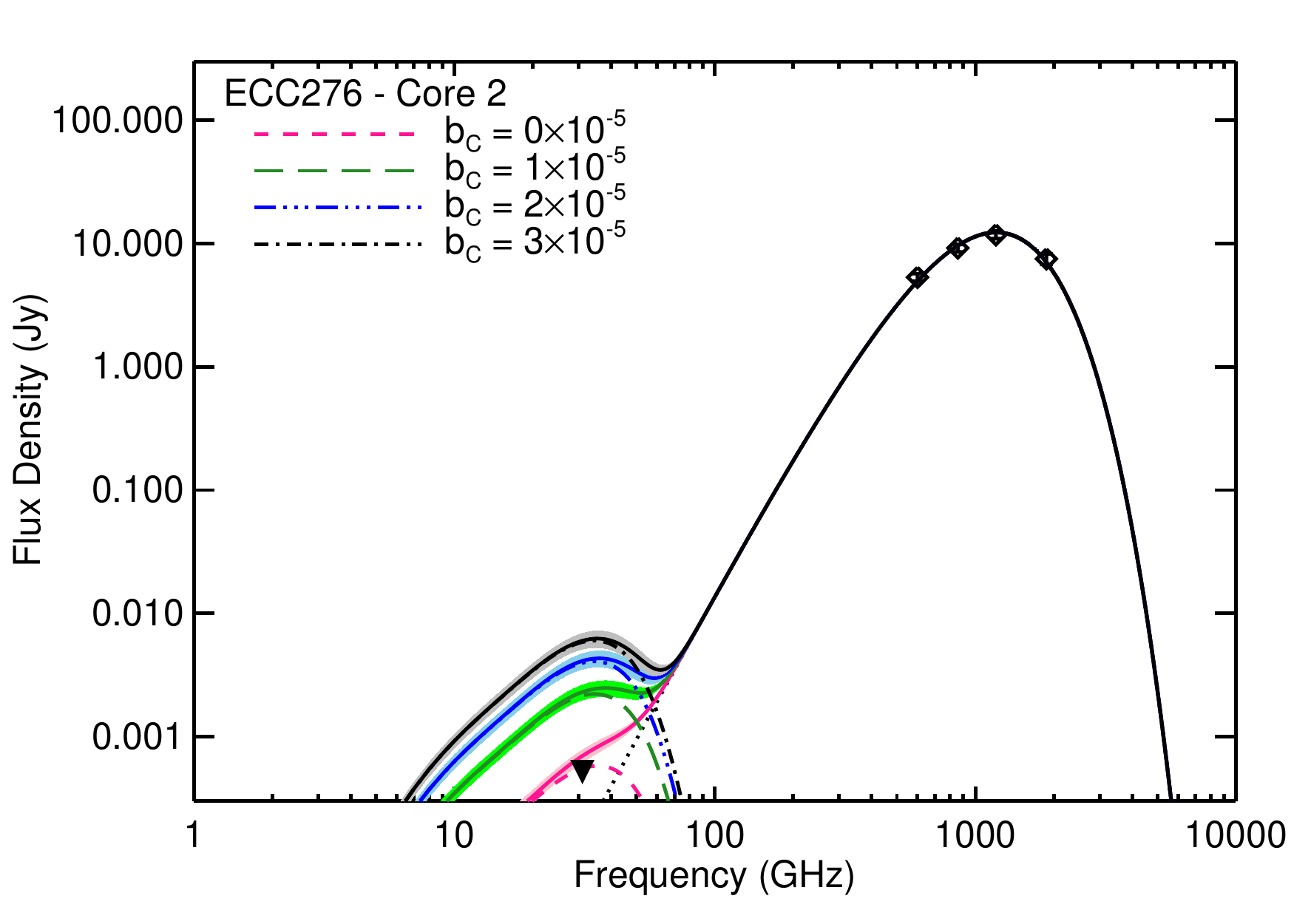} \\
\vspace{-0.35cm}
\includegraphics[angle=0,scale=0.450]{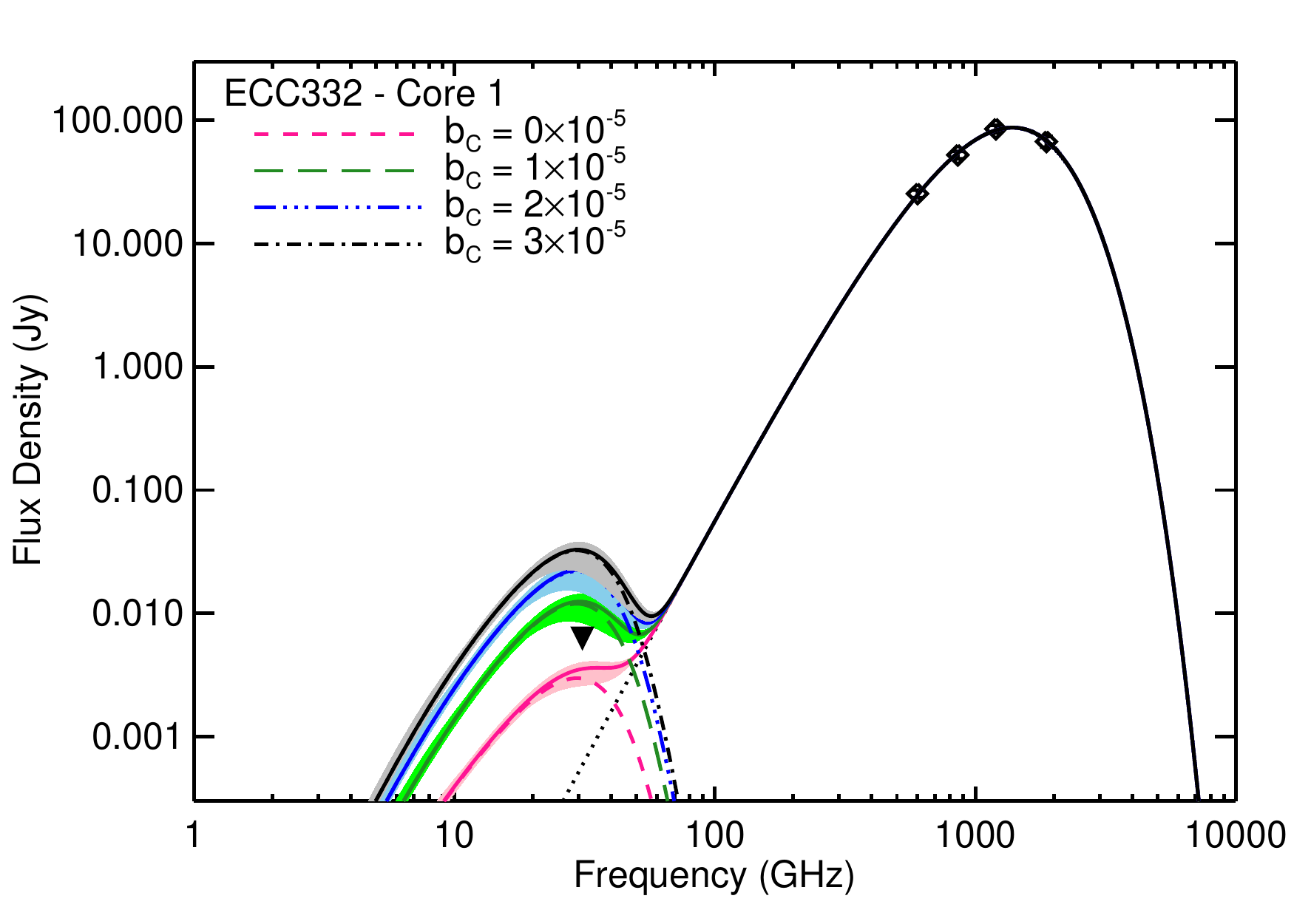}
\includegraphics[angle=0,scale=0.450]{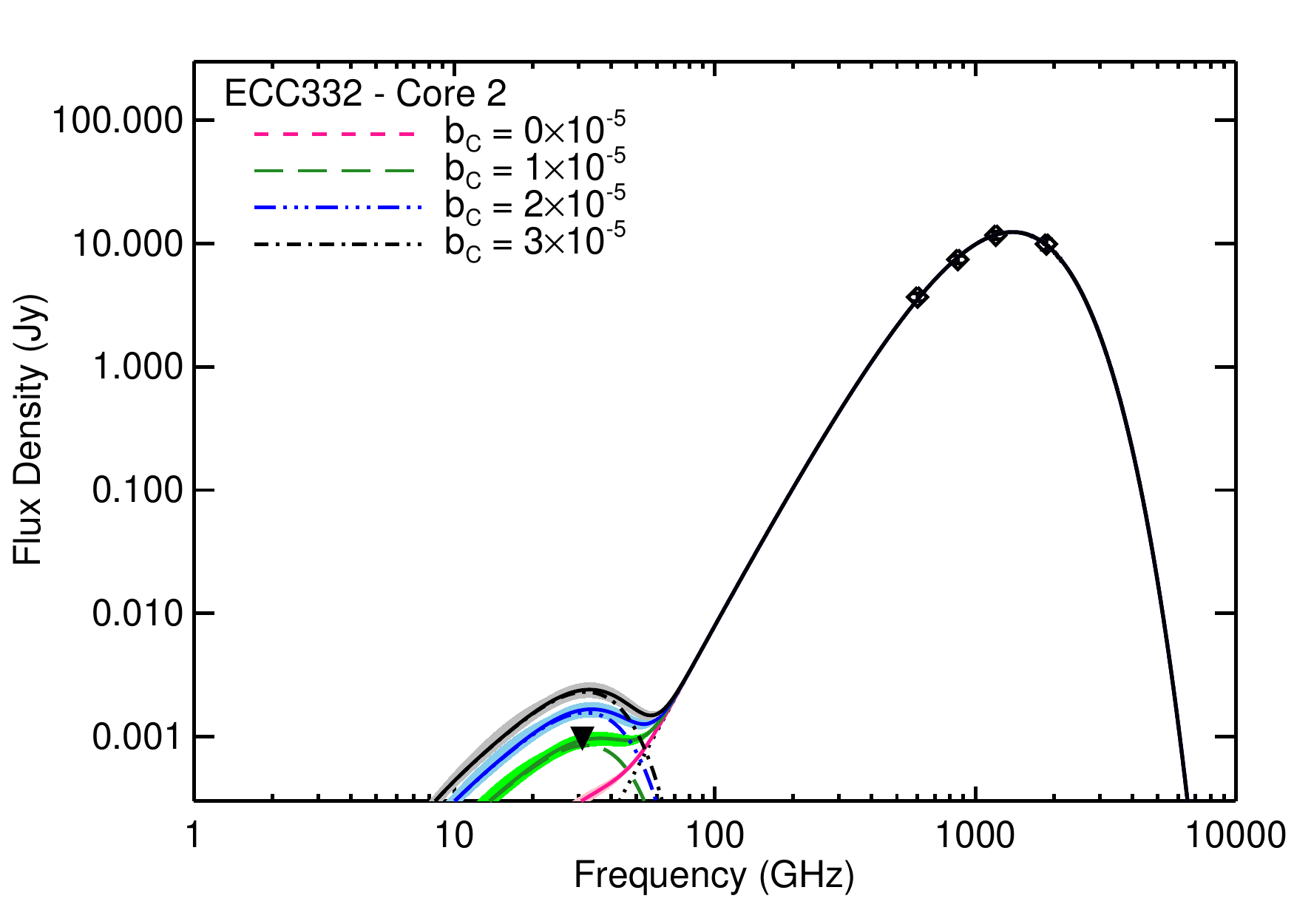} \\
\vspace{-0.35cm}
\includegraphics[angle=0,scale=0.450]{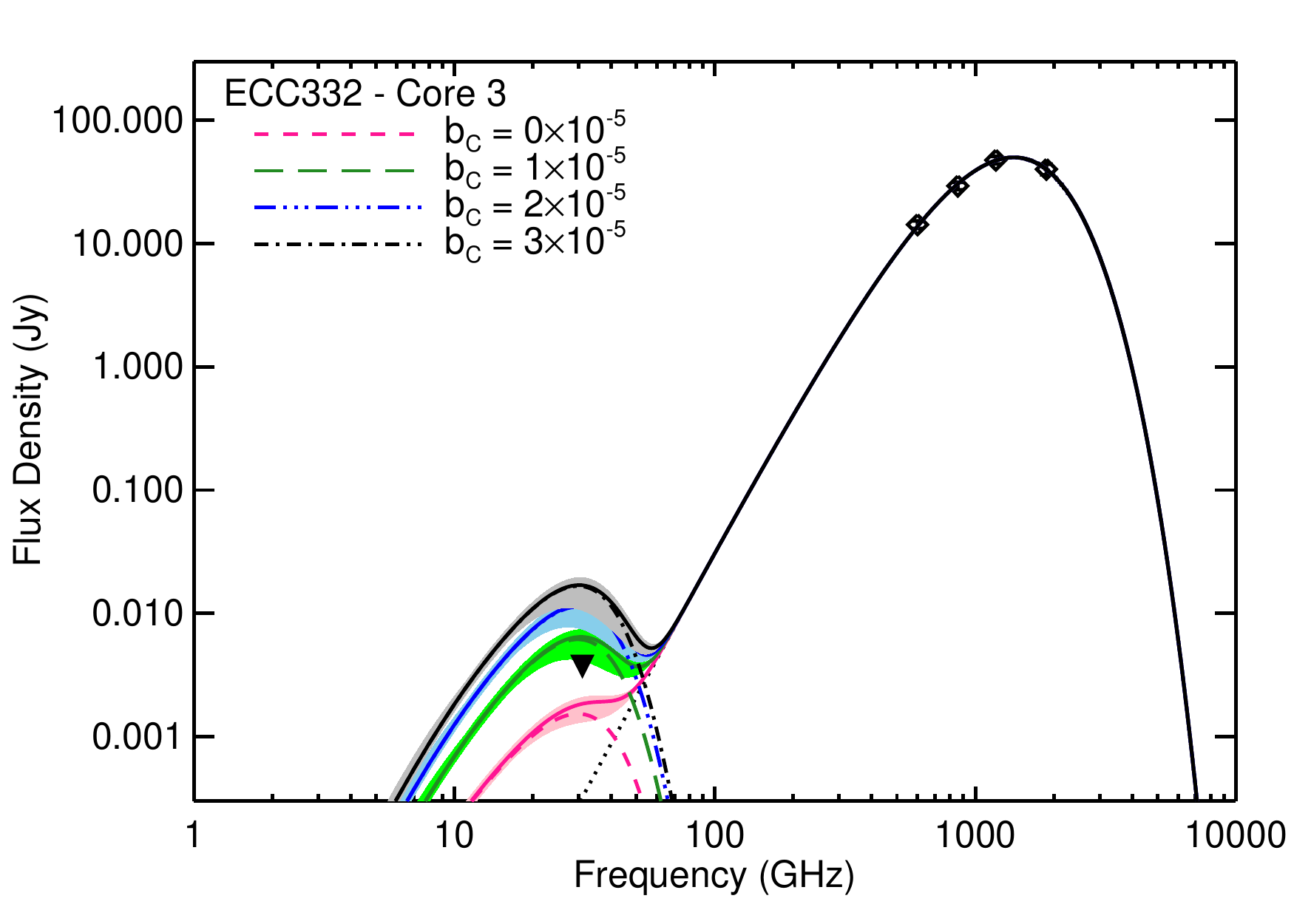} 
\includegraphics[angle=0,scale=0.450]{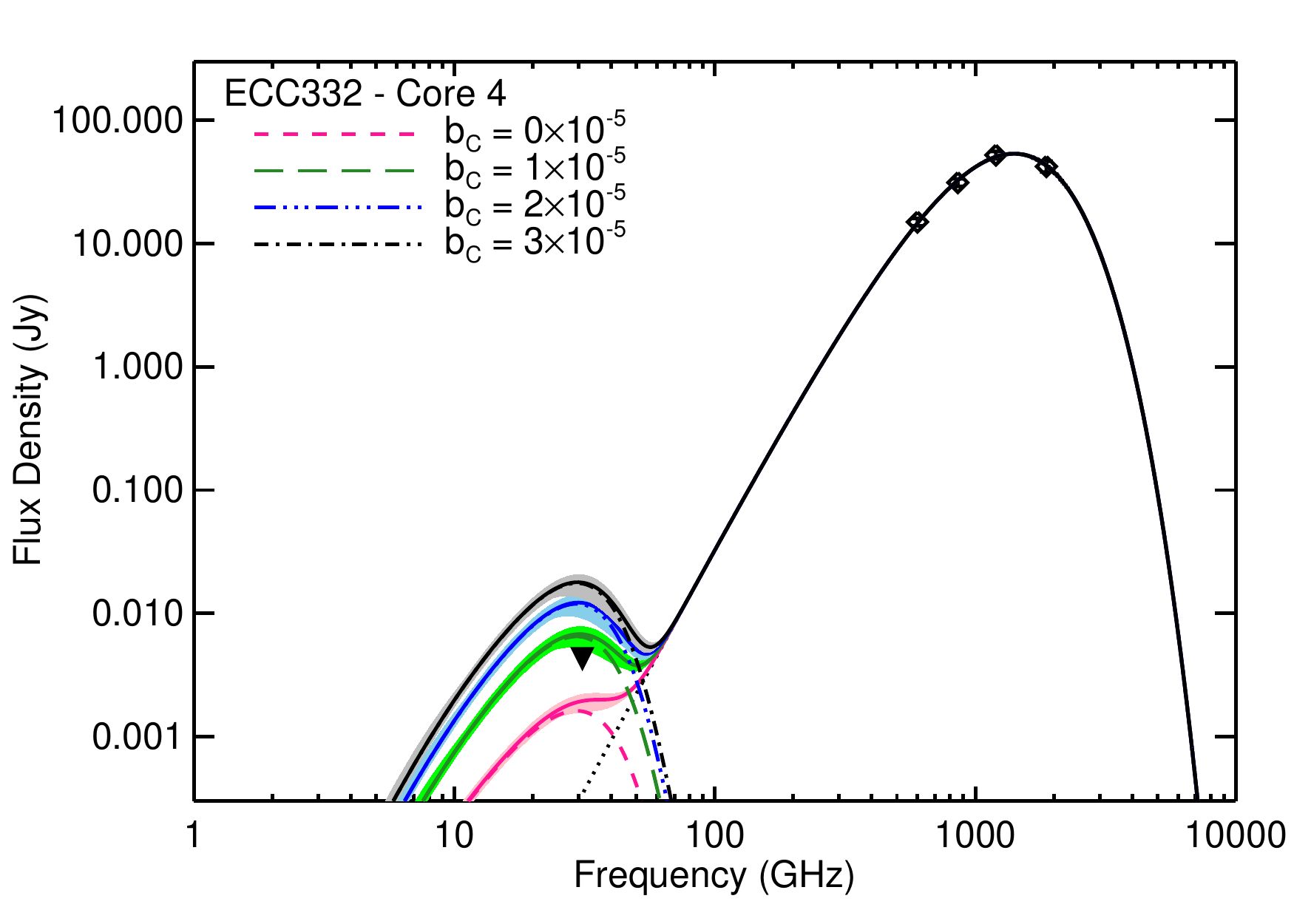} \\
\vspace{-0.35cm}
\includegraphics[angle=0,scale=0.450]{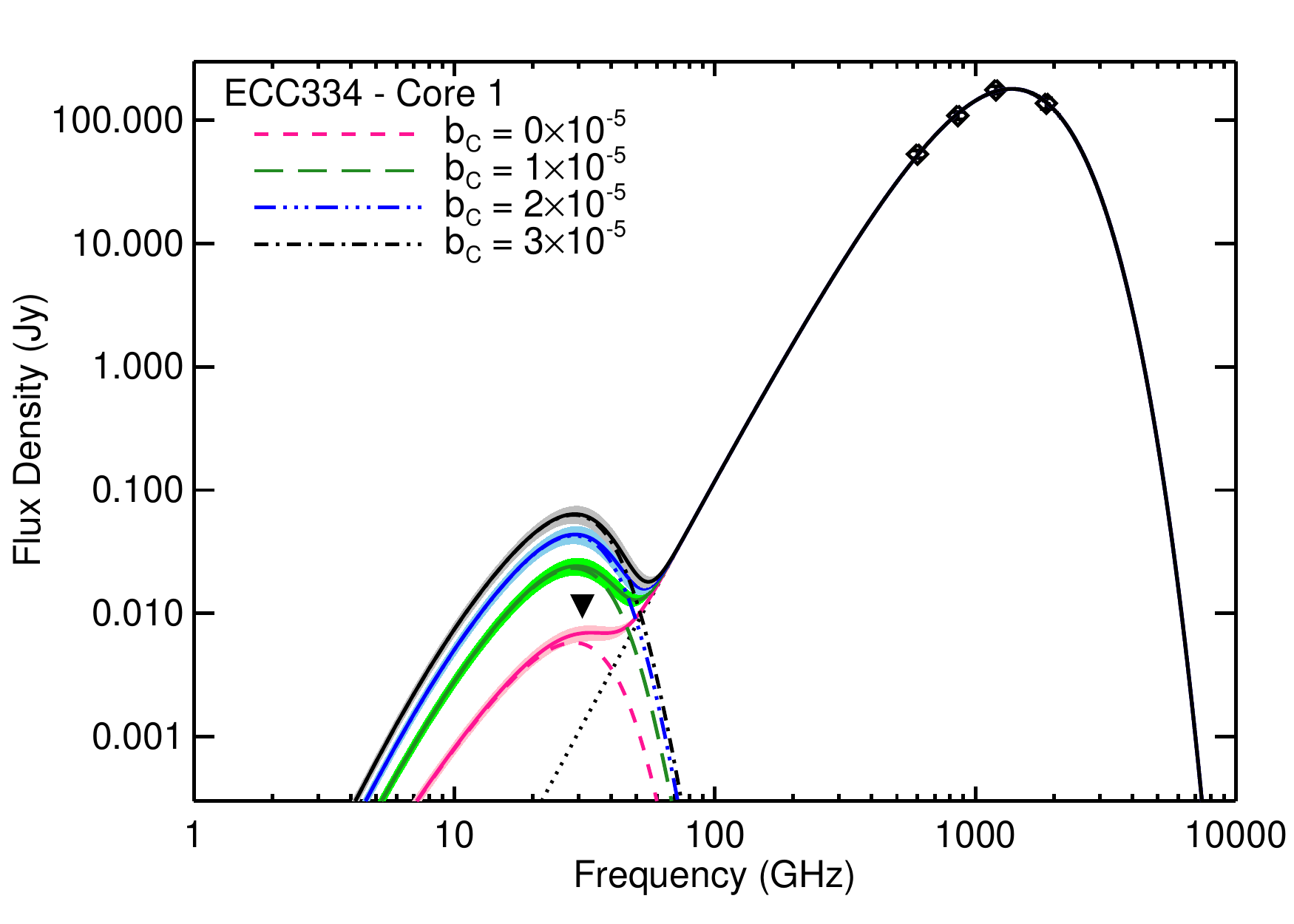}
\includegraphics[angle=0,scale=0.450]{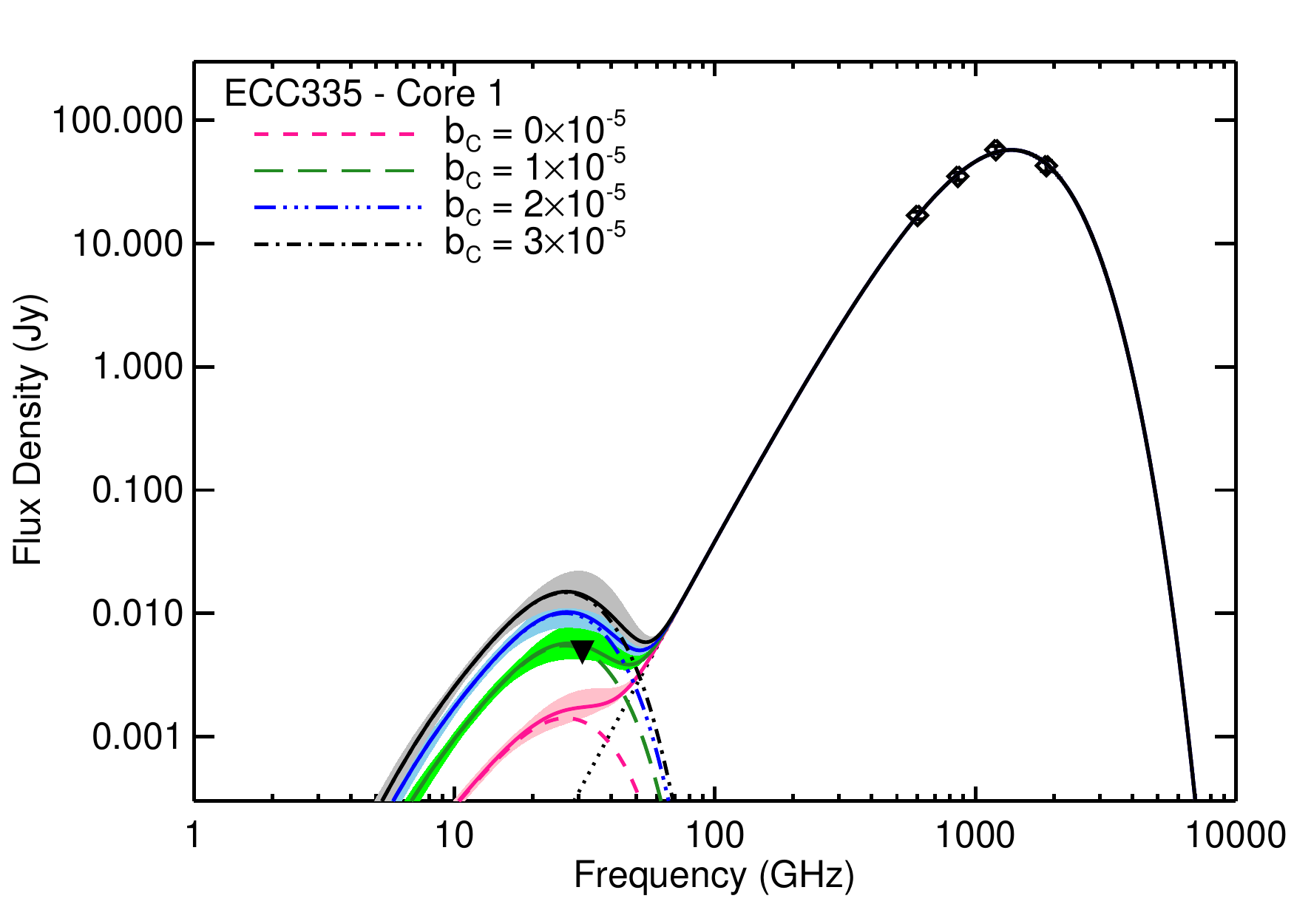} \\
\end{center}
\vspace{-0.30cm}
\caption{Continued}
\label{Fig:SED_sub_clumps}
\end{figure*}

\begin{figure*}
\ContinuedFloat
\begin{center}
\includegraphics[angle=0,scale=0.450]{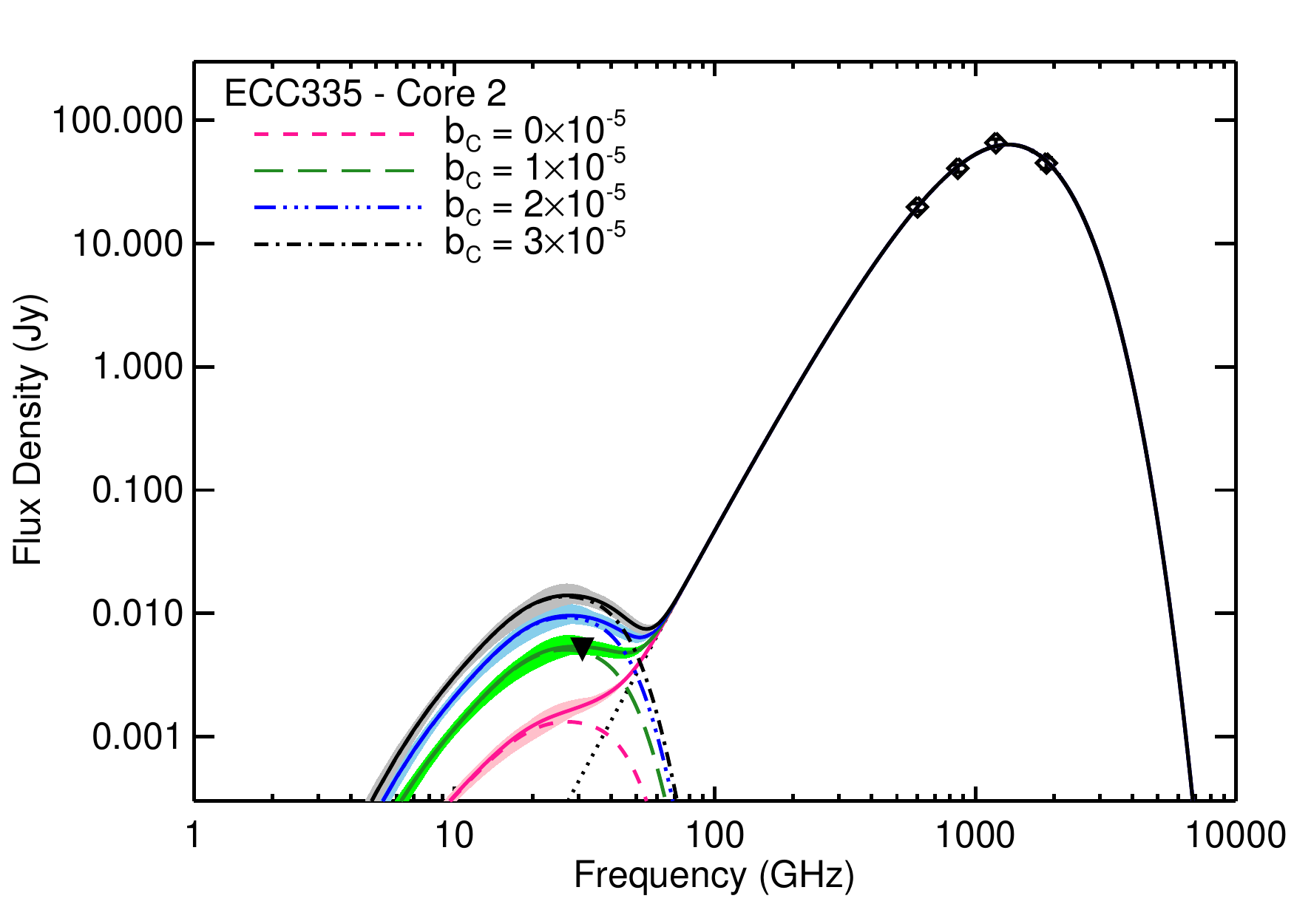}
\includegraphics[angle=0,scale=0.450]{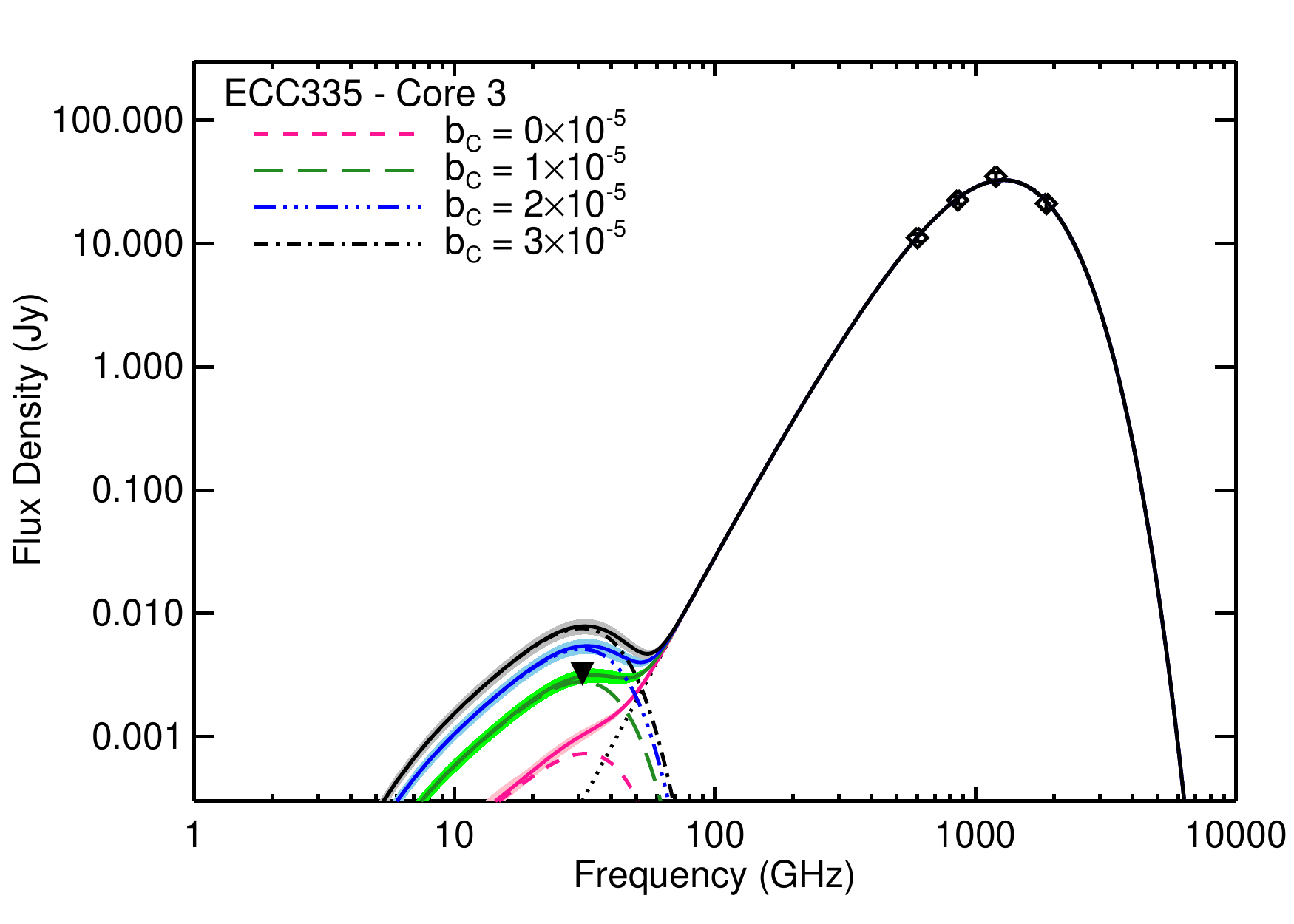} \\
\vspace{-0.35cm}
\includegraphics[angle=0,scale=0.450]{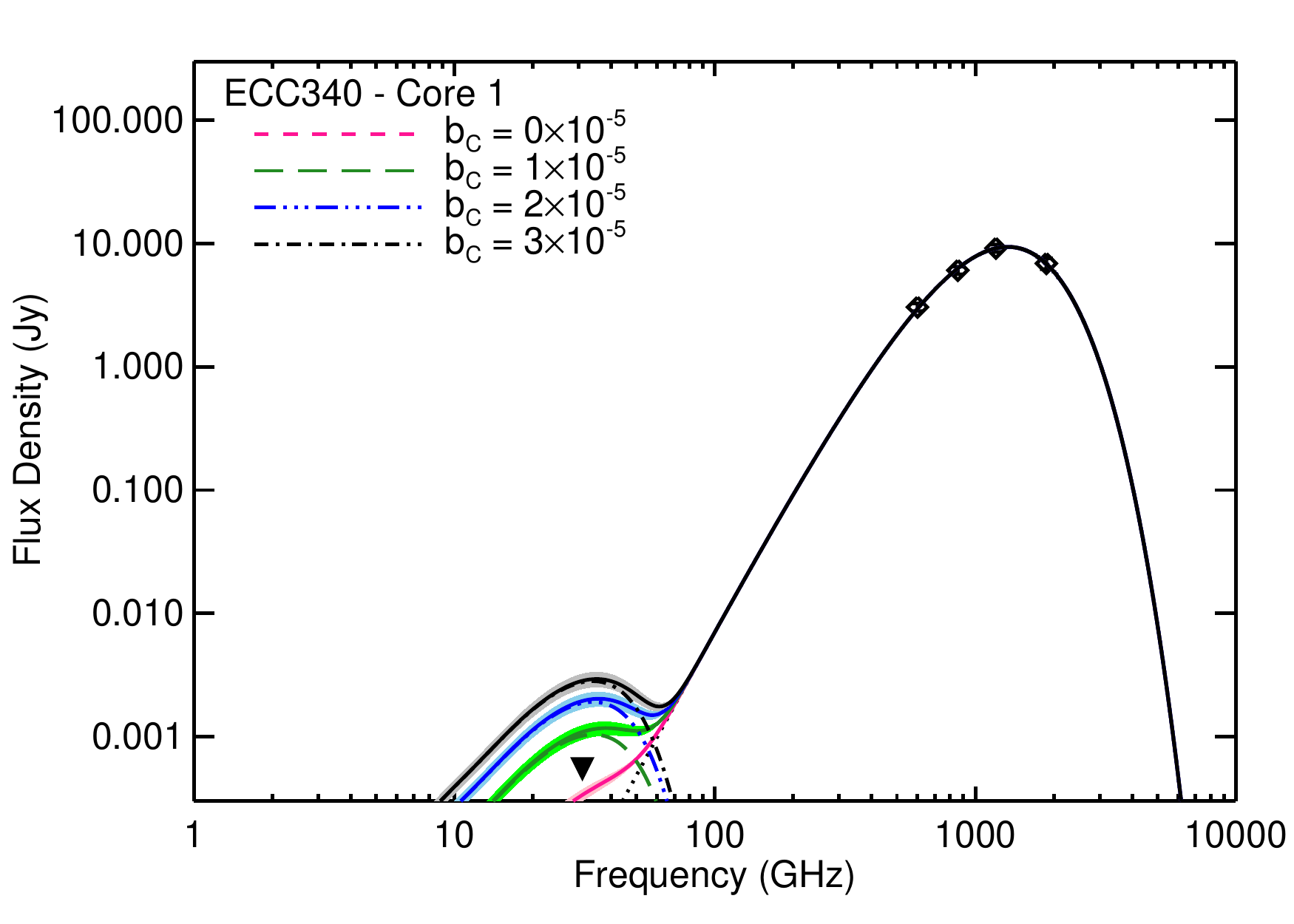}
\includegraphics[angle=0,scale=0.450]{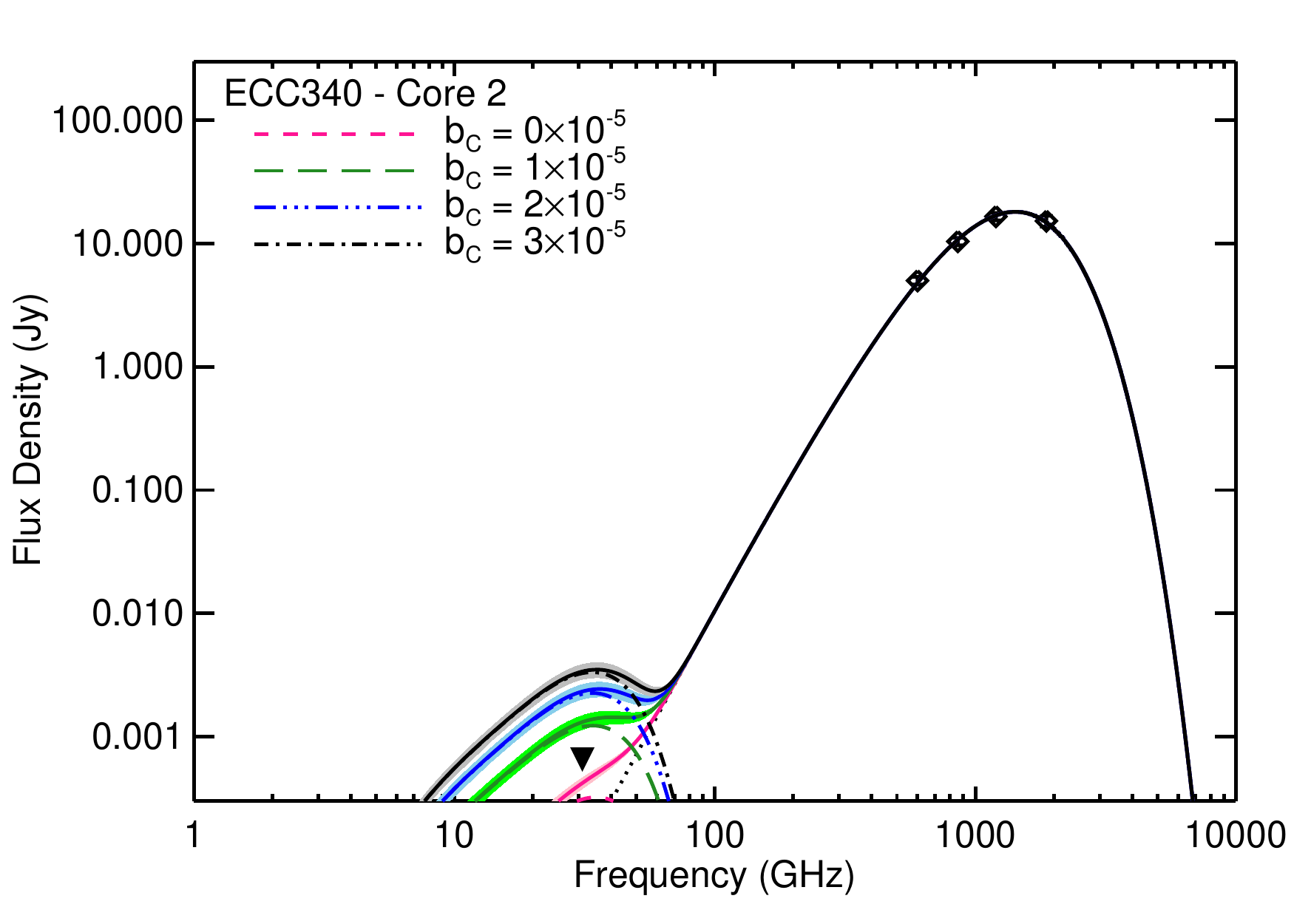} \\
\vspace{-0.35cm}
\includegraphics[angle=0,scale=0.450]{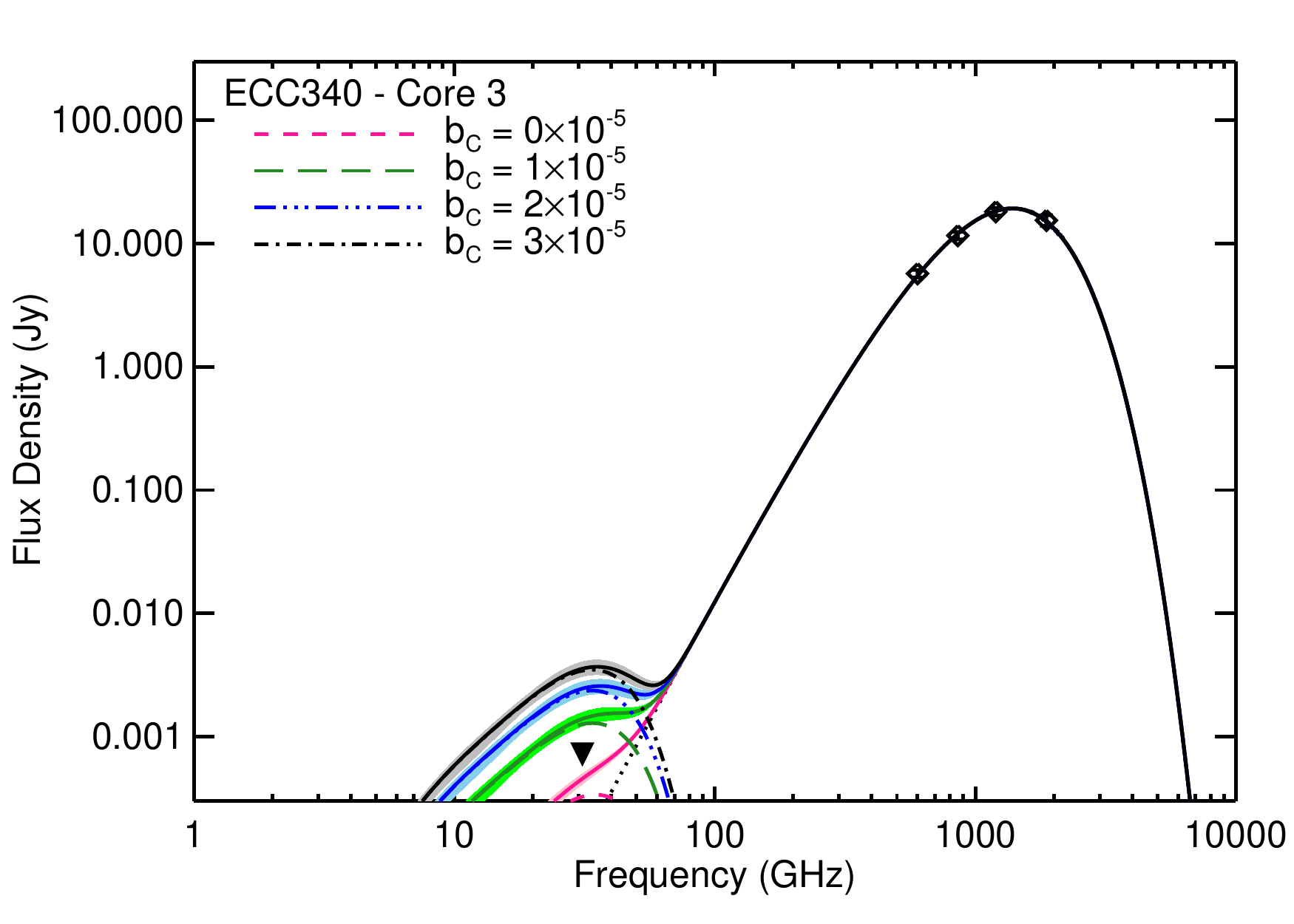}
\includegraphics[angle=0,scale=0.450]{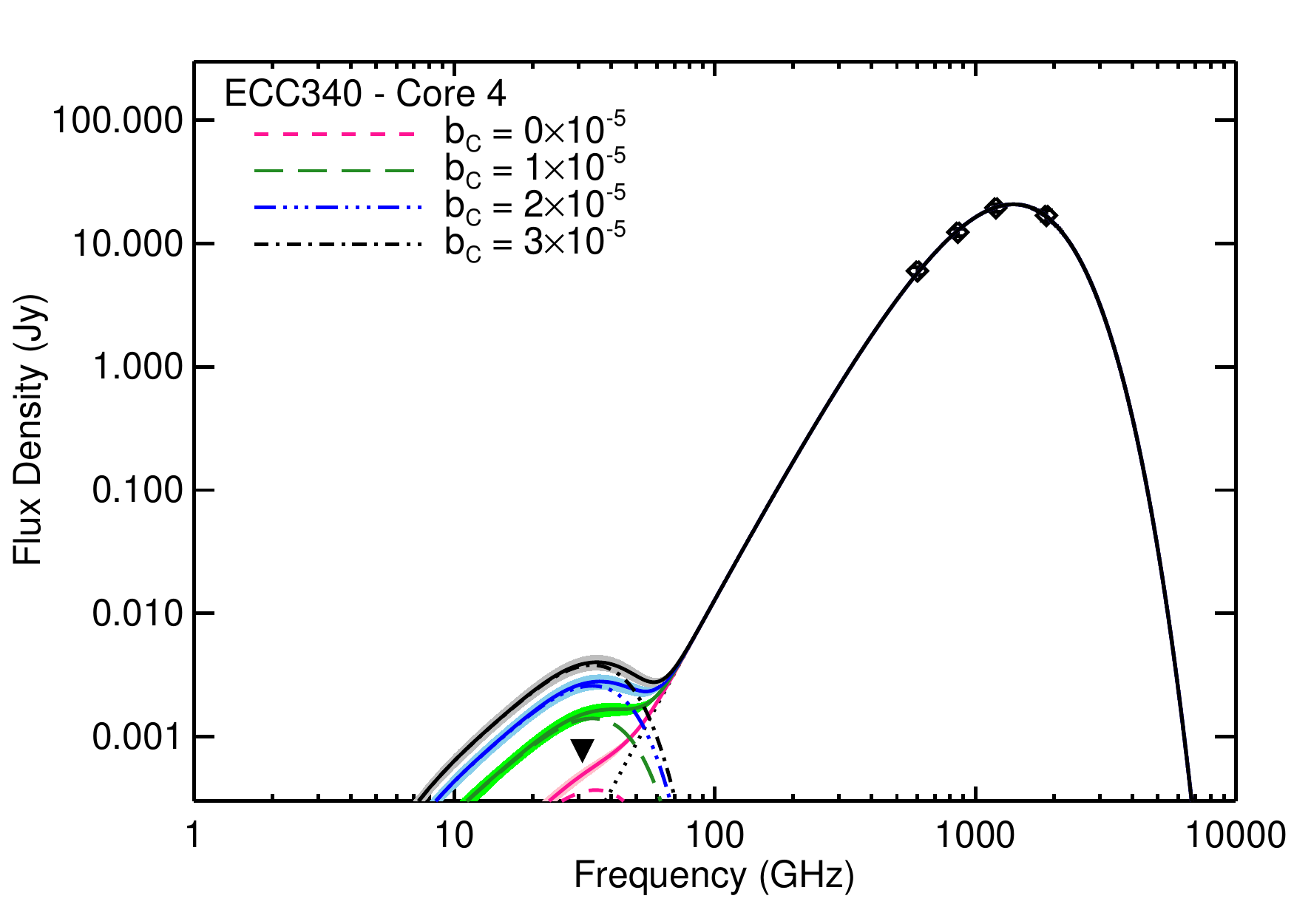} \\
\vspace{-0.35cm}
\includegraphics[angle=0,scale=0.450]{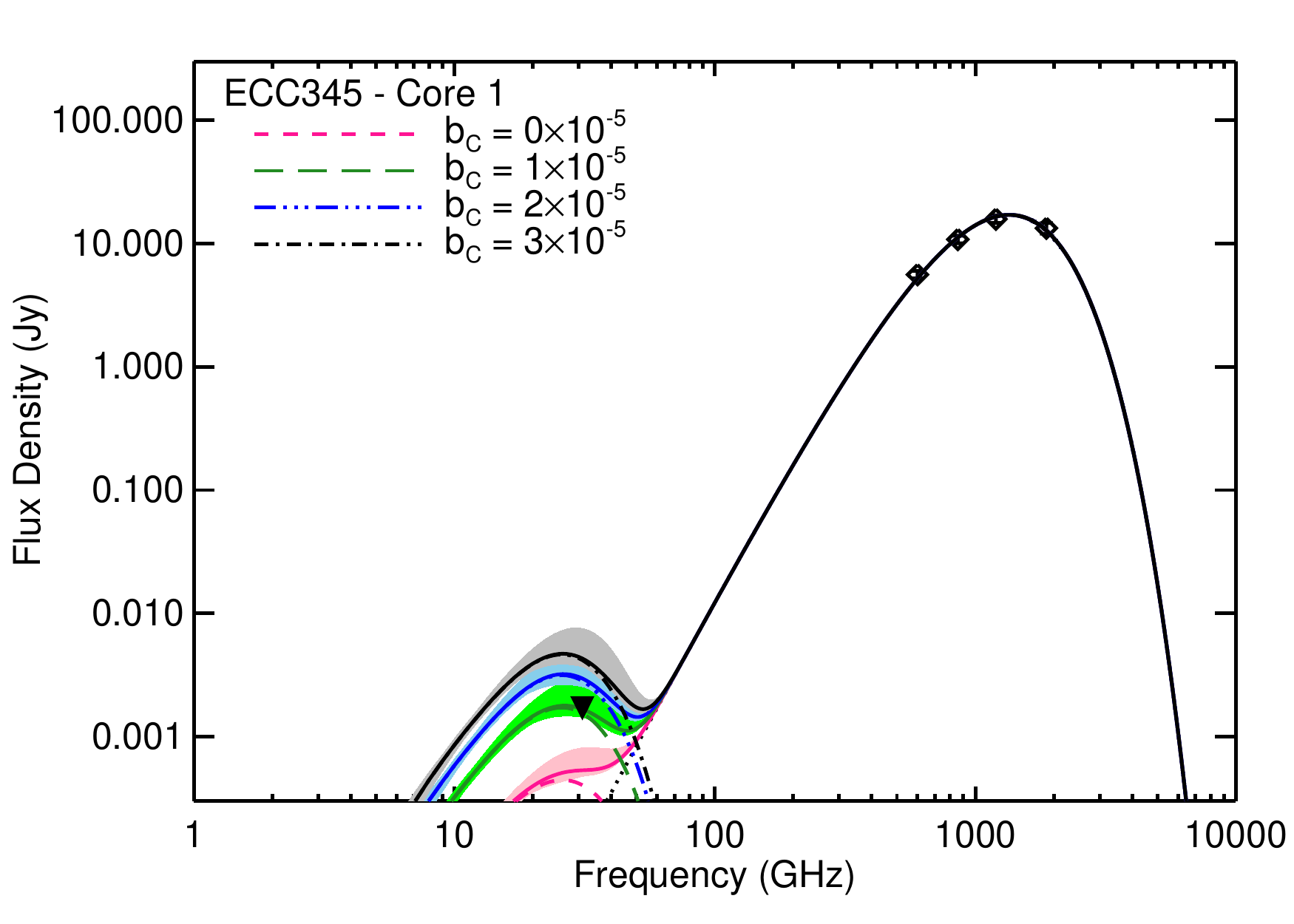}
\includegraphics[angle=0,scale=0.450]{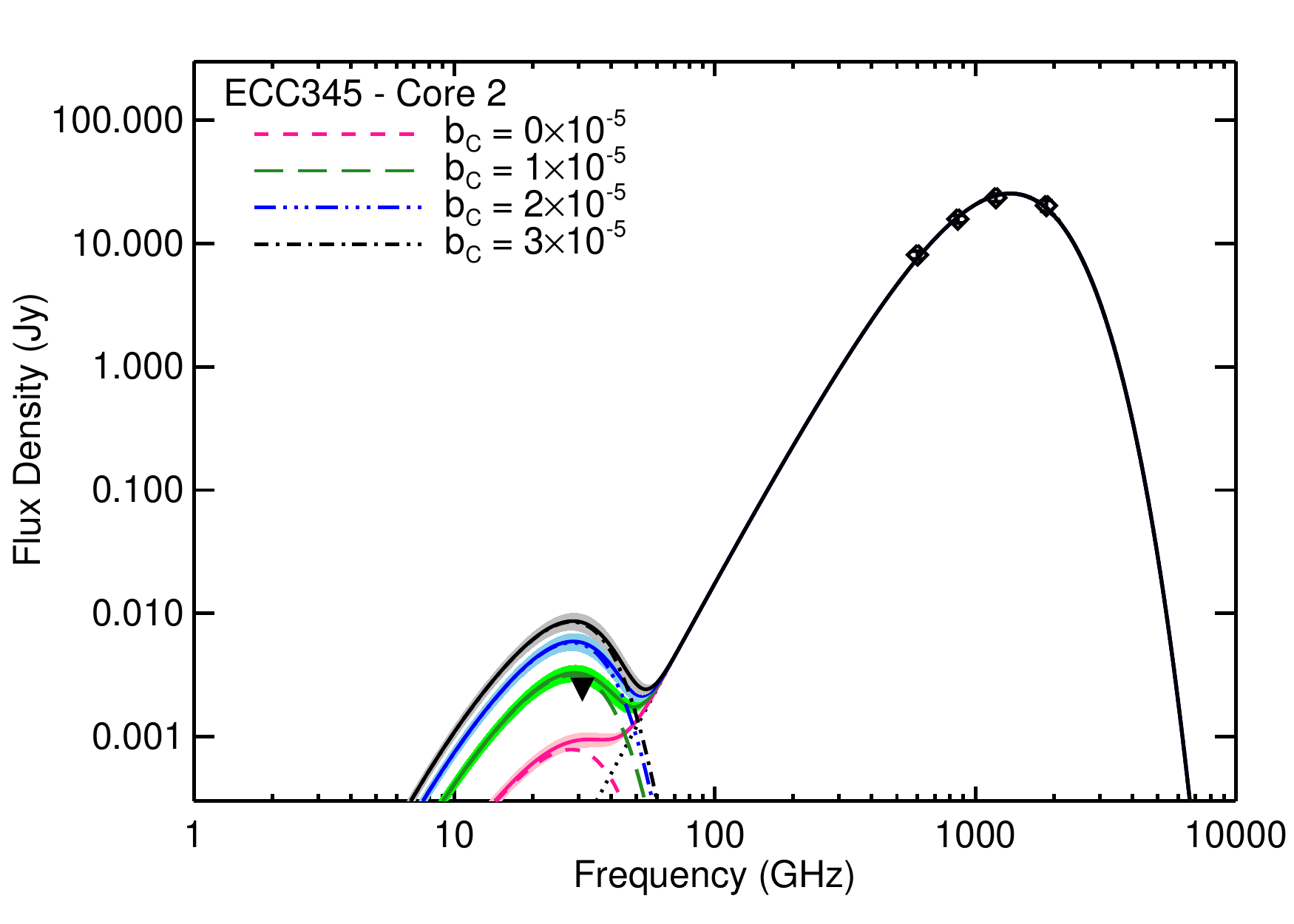} \\
\end{center}
\vspace{-0.30cm}
\caption{Continued}
\label{Fig:SED_sub_clumps}
\end{figure*}

\begin{figure*}
\ContinuedFloat
\begin{center}
\includegraphics[angle=0,scale=0.450]{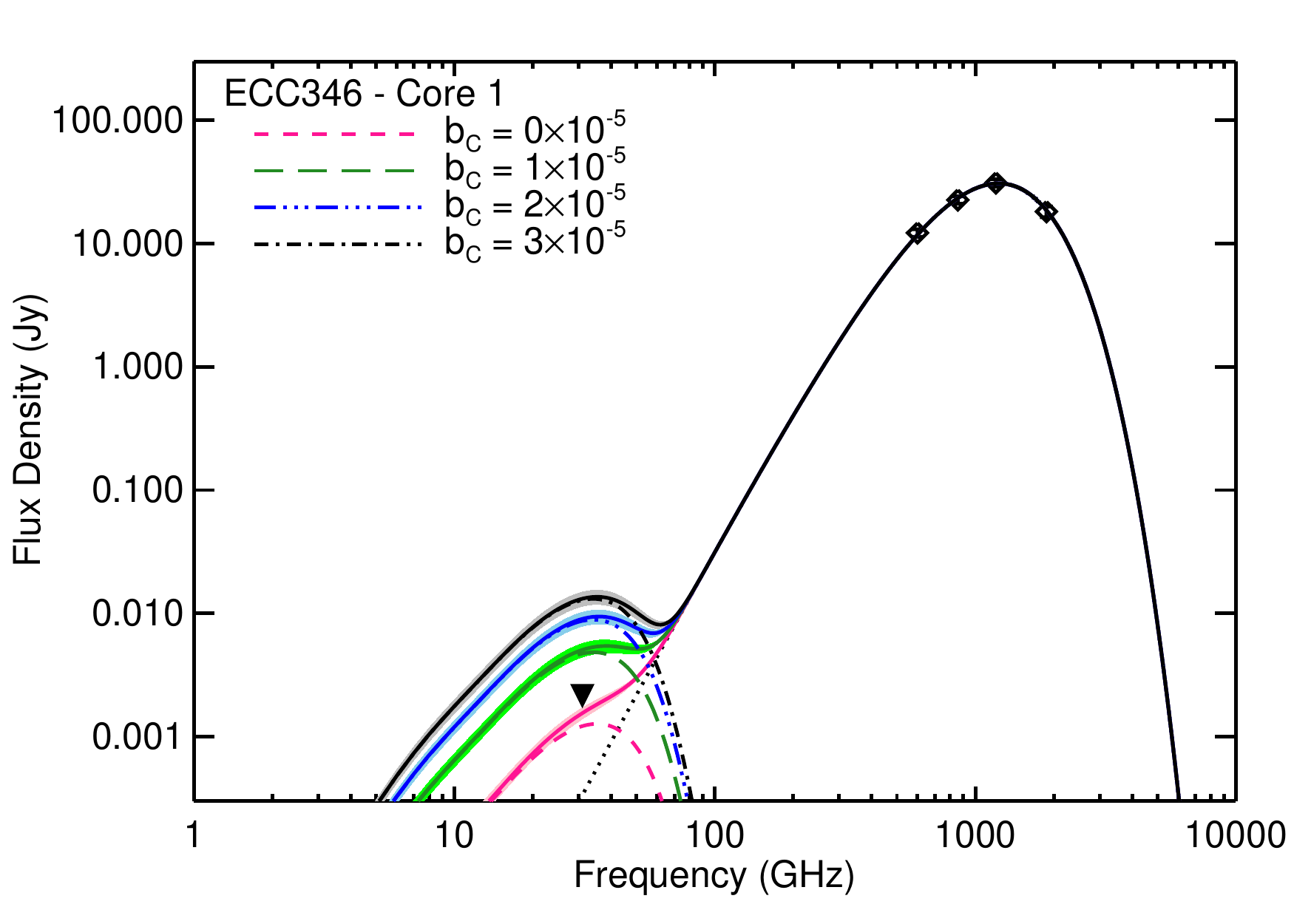}
\includegraphics[angle=0,scale=0.450]{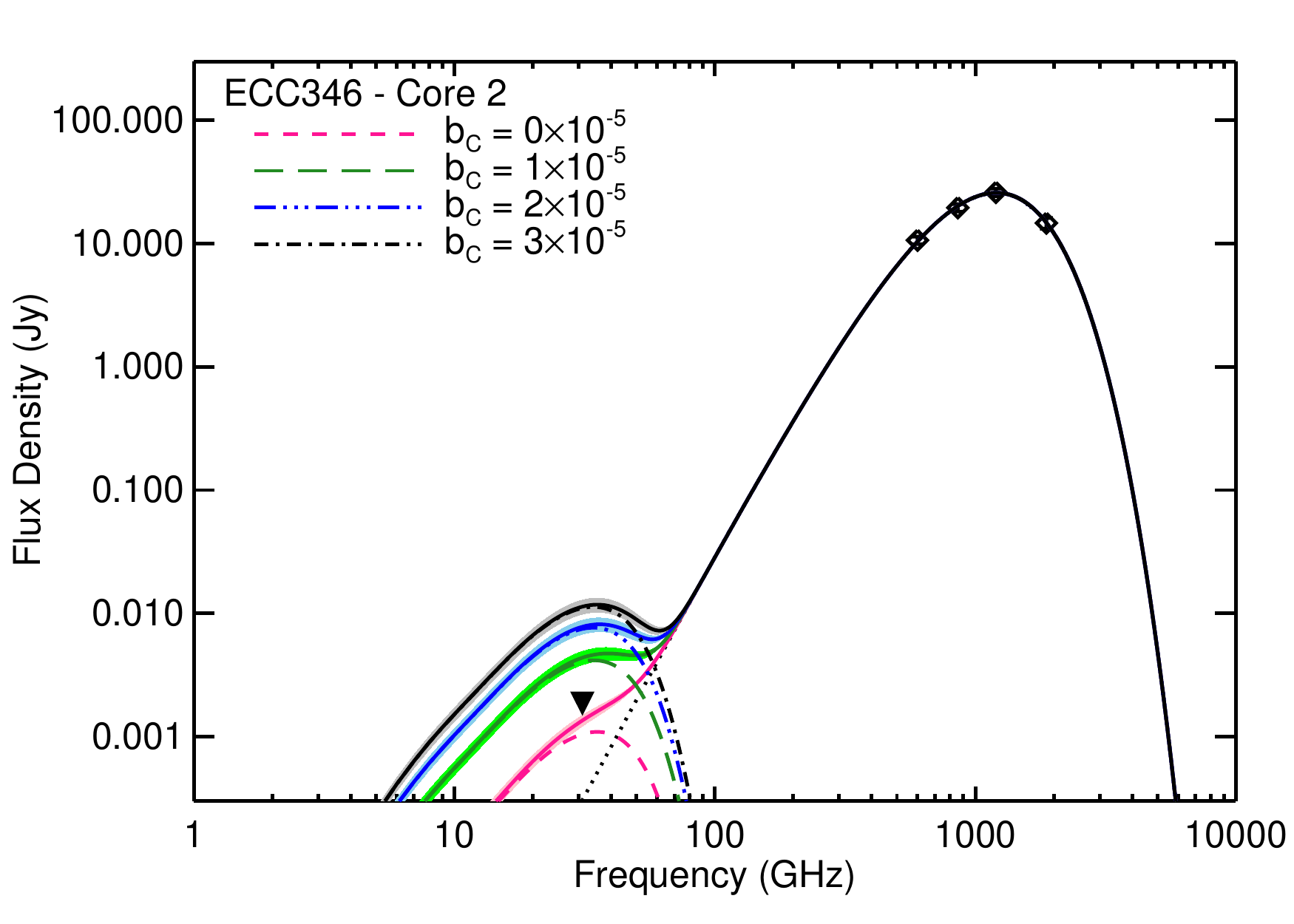}
\end{center}
\vspace{-0.30cm}
\caption{Continued}
\label{Fig:SED_sub_clumps}
\end{figure*}

%%%%%%% Modelling The Spinning Dust %%%%%%%%%%%%%%%%%%%%%%%%%%%%%%%%%%%%%%%%%%%%

\subsection{Modelling The Spinning Dust Emission}
\label{Subsec:Model_Spin_Dust}

In addition to the dust grain size distribution, the environmental conditions play an important role in modelling the spinning dust emission. Within \textsc{spdust}, the environmental conditions are parameterized using the following set of parameters: the density, $n_{\mathrm{H}}$; the radiation field, $G_{\mathrm{0}}$; the gas temperature, $T_{\mathrm{gas}}$; the hydrogen ionization fraction, $x_{\mathrm{H}}$; the carbon ionization fraction, $x_{\mathrm{C}}$; and the molecular hydrogen fraction, $y$. Since~\citet{Tibbs:15a} used \textit{Herschel} data to derive values for $\bar{n}$$_{\mathrm{H}}$ and $\bar{G}$$_{\mathrm{0}}$ for all of the cores~(see Table~\ref{Table:Clump_Props}), and we have adopted the values of $T_{\mathrm{gas}}$ derived by~\citet{Wu:12}~(see Table~\ref{Table:Sources}), we just need to determine appropriate values for the remaining parameters~($x_{\mathrm{H}}$, $x_{\mathrm{C}}$, and $y$). For these parameters we use the values defined by~\citet{DaL:98} for dark clouds. Given the densities and dust temperatures of our cores, the central regions will be shielded from the interstellar radiation field~(we find that $\bar{G}$$_{\mathrm{0}} << 1$) and so there will be very few ionizing photons available to ionize the gas (i.e., $x_{\mathrm{H}}$~=~0 and $x_{\mathrm{C}}$~=~10$^{-6}$) or to photo-dissociate molecular hydrogen~(i.e., $y$~=~0.999).

\begin{table}
\begin{center}
\caption{Parameters used to model the spinning dust emission in our cores.}
\begin{tabular}{ll}
\hline
 &  \\
 &  \\
\hline
\hline

$n_{\mathrm{H}}$ ($\mathrm{H~cm}^{-3}$)                            			& See Table~\ref{Table:Clump_Props} \\
$G_{\mathrm{0}}$ 											& See Table~\ref{Table:Clump_Props} \\
$T_{\mathrm{gas}}$ (K)                   								& See Table~\ref{Table:Sources} \\
$x_{\mathrm{H}}$ $\equiv$ $n_{\mathrm{H^{+}}}$/$n_{\mathrm{H}}$ 	& 0 \\
$x_{\mathrm{C}}$ $\equiv$ $n_{\mathrm{C^{+}}}$/$n_{\mathrm{H}}$ 	& 10$^{-6}$ \\
$y$ $\equiv$ 2$n_{\mathrm{H_{2}}}$/$n_{\mathrm{H}}$ 				& 0.999 \\
$\mu$ $\mid_{\mathrm{a = 1}\mathrm{nm}}$ (Debye) 				& 9.3 \\

\hline
\label{Table:Spin_Params}
\end{tabular}
\end{center}
\textbf{Notes}: $n_{\mathrm{H}}$ is the hydrogen number density, $G_{\mathrm{0}}$ is the radiation field, where a value of 1 corresponds to the~\citet{Mathis:83} solar neighbourhood radiation field, $T_{\mathrm{gas}}$ is the gas temperature, $x_{\mathrm{H}}$ is the hydrogen ionization fraction, $x_{\mathrm{C}}$ is the carbon ionization fraction, $y$ is the molecular hydrogen fraction, and \textbf{$\mu$ $\mid_{\mathrm{a = 1}\mathrm{nm}}$} is the rms dipole moment. The values for $n_{\mathrm{H}}$ and $G_{\mathrm{0}}$ are the values for each core listed in Table~\ref{Table:Clump_Props}, and the values of $T_{\mathrm{gas}}$ are listed in Table~\ref{Table:Sources}.
\end{table}

There is one additional parameter that is required to model the spinning dust emission, the electric dipole moment, $\mu$. This is one of the most uncertain parameters in the spinning dust model. We assume that the rms dipole moment for a grain of size $a$~=~1~nm is 9.3~D, which is equivalent to an average rms dipole moment per atom of 0.38~D. Although this value is highly uncertain, it is in agreement with both the original value adopted by~\citet{DaL:98}, who used 0.4~D based on laboratory measurements, and with the observational constraints placed by~\citet{Ysard:10a}, who found a range between 0.3~--~0.4~D. We note that larger dipole moments may also be possible. For example, if regular polycyclic aromatic hydrocarbons~(PAHs) incorporate a nitrogen atom within their aromatic carbon skeleton, they will become polycyclic aromatic nitrogen heterocycles~(PANHs), which have dipole moments of~$\sim$1.5~--~10~D, corresponding to~$\sim$0.6~--~0.7~D per atom~\citep{Hudgins:05}. In fact, this increase in the observed dipole moment led \citet{Hudgins:05} to propose that these PANHs could be the carriers of the spinning dust emission. Besides PANHs, recent dust models~\citep{Kwok:13, Jones:13} have implemented a more general mixture of aromatic and aliphatic amorphous carbon to replace PAHs, based on the evidence suggesting that PAHs are not the dominant hydrocarbon component in the ISM. Such amorphous carbon molecules are highly asymmetric and three dimensional, which as discussed by~\citet{Hoang:11} results in an increase in the spinning dust emissivity.

In Table~\ref{Table:Spin_Params} we list the complete set of input parameters that we use to model the spinning dust emission.

%%%%%%% Abundance And Coagulation %%%%%%%%%%%%%%%%%%%%%%%%%%%%%%%%%%%%%%%%%%%%

\section{Discussion}
\label{sec:Abundance_Coagulation}

%%%%%%% Abundance  %%%%%%%%%%%%%%%%%%%%%%%%%%%%%%%%%%%%%%%%%%%%

\subsection{Constraining the Abundance of Very Small Grains}
\label{Subsec:Constrain_Abundance}

To constrain the abundance of very small grains in all of the cores, the spinning dust emission was modelled, as described in Section~\ref{Subsec:Model_Spin_Dust}, for the four different values of $b_{\mathrm{C}}$ discussed in Section~\ref{Subsec:Size_Dist}. To incorporate the uncertainty on $\bar{n}$$_{\mathrm{H}}$ and $\bar{G}$$_{\mathrm{0}}$, we repeated the modelling analysis 1000 times, each time randomising the input $\bar{n}$$_{\mathrm{H}}$ and $\bar{G}$$_{\mathrm{0}}$ values within their uncertainty. The resulting spinning dust curves, along with the thermal dust emission curves, are displayed in Figure~\ref{Fig:SED_sub_clumps}. These plots illustrate how strongly the spinning dust emission depends on the abundance of very small grains as it is clear to see that increasing $b_{\mathrm{C}}$ from 0$\times$10$^{-5}$ to 3$\times$10$^{-5}$ increases the amplitude of the spinning dust emission i.e., an increase in the very small dust grain abundance results in an increase in the strength of the spinning dust emission. The thermal dust emission in Figure~\ref{Fig:SED_sub_clumps} is modelled using a modified black body function fitted to the \textit{Herschel} flux densities at 160, 250, 350, and 500~$\mu$m, with a fixed dust opacity index of $\beta$~=~2. As discussed by~\citet{Tibbs:15a}, this is appropriate for these dense environments, however, we emphasize that using a flatter value of $\beta$~=~1.8 does not significantly change the results of our analysis. 

By comparing the total modelled emission~(spinning dust $+$ thermal dust) at 1~cm with the observed 1~cm emission, we can determine the abundance of very small grains in these cores. Therefore, we also plotted the~\citet{Tibbs:15a} CARMA 1~cm measurements~($S_{1~\mathrm{cm}}^{\mathrm{observed}}$) on the SEDs in Figure~\ref{Fig:SED_sub_clumps}. These flux densities are listed in Table~\ref{Table:Clump_Props}, and as discussed by~\citet[][]{Tibbs:15a}, there is only a detection for three of the cores, with the remaining cores being limited to a conservative 5$\sigma$ upper limit. Even with just a conservative upper limit, when we compare the 1~cm CARMA data with the predicted spinning dust curves in Figure~\ref{Fig:SED_sub_clumps}, we find that we can still place a constraint on the very small dust grain abundance. The values of $b_{\mathrm{C}}$ that are consistent with the CARMA data are listed in Table~\ref{Table:Clump_Props}. For all of our cores we find that $b_{\mathrm{C}}$~$\le$~1$\times$10$^{-5}$, in agreement with~\citet{Tibbs:15a}, who concluded that the observed cm emission was below the predicted spinning dust emission for all of the cores for $b_{\mathrm{C}}$~$=$~3$\times$10$^{-5}$. For ECC229 cores 1 and 3 we find that $b_{\mathrm{C}}$~=~0$\times$10$^{-5}$. This does not imply that the total abundance of very small grains is zero, but it actually implies that there is no contribution from the two log-normal components in the dust grain size distribution i.e., $D(a)$ = 0 in Equation~\eqref{equ:dnda}. From Figure~\ref{Fig:Size_Dist} it is clear to see that even with $b_{\mathrm{C}}$~=~0$\times$10$^{-5}$, there is still a tail to the size distribution that extends to very small grain radii. Additionally, for ECC229 core 4, ECC276 core 1, and ECC276 core 2 we find that $b_{\mathrm{C}}$~$<$~0$\times$10$^{-5}$, which implies that even with $D(a)$~=~0, the size distribution described by Equation~\eqref{equ:dnda} over-predicts the abundance of very small grains.

%%%%%%% Coagulation %%%%%%%%%%%%%%%%%%%%%%%%%%%%%%%%%%%%%%%%%%%%

\subsection{Deficit Of Very Small Grains} 
\label{Subsec:Deficit}

For the diffuse Galactic ISM~($R_{\mathrm{V}}$~=~3.1), \citet{Li:01} found that the observed IR emission and extinction were best fitted with a value of $b_{\mathrm{C}}$ = 6$\times$10$^{-5}$. This value is larger than the values we estimated for our sample of cores, implying that we have detected a deficit in the abundance of very small dust grains in these dense environments relative to the diffuse ISM.

A decrease in the abundance of very small grains in dense environments has previously been observed~\citep[e.g.,][]{Laureijs:91, Kramer:03, Stepnik:03, Ysard:13}, however, these previous works all based their analysis on data at IR/sub-mm wavelengths. Our analysis represents the first detection of a deficit of very small grains in a dense environment using cm observations of spinning dust emission. The decrease in the very small dust grain abundance has previously been attributed to dust grain evolution through the process of grain growth, but can that account for our results? 

Grain growth can occur via the accretion of metals from the gas phase onto the grains, and/or through grain-grain coagulation. Dust grains of all sizes can grow via accretion, however, since the very small grains contribute a larger surface area than the big grains, it is the very small grains that accrete most and hence grow in size more rapidly than the big grains. Like for accretion, the process of grain-grain coagulation involves dust grains of all sizes. Coagulation can occur between any two dust grains regardless of grain size (e.g., very small grain onto very small grain, very small grain onto big grain, and big grain onto big grain), however, simulations show that very small grains coagulate faster than big grains~\citep{Ossenkopf:93}. \citet{Kohler:12} estimated the timescales for such coagulation processes and found that the coagulation of very small grains onto big grains had the shortest timescales~($\sim$1.6$\times$10$^{3}$ years), approximately 2 orders of magnitude faster than the coagulation of very small grains onto very small grains or big grains onto big grains. The accretion timescale is always short compared to the coagulation timescale~\citep{Ossenkopf:93, Jones:13}, and hence accretion enhances the influence of coagulation by increasing the collision cross-sections and sticking coefficients of the grains.

A lower limit on the lifetime of cold cores is given by the free-fall timescale, $\tau_{\mathrm{ff}}$,

\begin{equation}
\tau_{\mathrm{ff}} = \left( \frac{3 \pi}{32 G \rho} \right) ^{1/2} ,
\end{equation}

\noindent
where $G$ is the gravitational constant and $\rho$ is the mass density of the core. For our cores this results in an age range of~$\sim$10$^{5}$~--~10$^{6}$~years. Since this is greater than the coagulation timescale, it implies that both accretion and coagulation are important processes in how the grains are growing in our cores, and it is likely that both accretion and coagulation occur. Such a grain growth process is efficient at removing the very small grains from the grain size distribution~\citep{Hirashita:12}. Therefore, given that theoretical models predict that grain growth can effectively decrease the abundance of very small grains, and that the timescale for such a process is less than the age of our sources, this suggests that grain growth via accretion and coagulation can explain the deficit of very small grains that we observe in these cold, dense cores.

In addition to a deficit of very small grains, grain growth should also give rise to an increase in the dust opacity at far-IR wavelengths, and in a recent analysis of a sample of Galactic cores,~\citet{Juvela:15} observed an increase in the dust opacity at 250~$\mu$m, which they interpreted as evidence for grain growth in dense molecular environments.

\subsection{Limitations Of This Analysis}
\label{Subsec:Limitations}

This analysis relies on the the assumption that the spinning dust hypothesis is the correct interpretation of the anomalous microwave emission i.e., that very small dust grains contain an electric dipole moment, and when spinning, produce electric dipole radiation. Furthermore, we rely on accurately modelling this spinning dust emission based on the physical properties of the cores, and as such, our results are limited by our model. We note that \textsc{spdust} does not incorporate some important effects such as internal thermal fluctuations, impulsive excitation due to high-impact ion collisions, and irregular dust grain shapes, all of which produce an increase in the emissivity of the spinning dust emission. Nevertheless, \textsc{spdust} is the only publicly available spinning dust model, and it has been shown to accurately fit observations of spinning dust emission~\citep[e.g.,][]{Tibbs:12a, Planck_Dickinson:14, Genova-Santos:15}.

In addition to these limitations of \textsc{spdust}, we are also dependent on the parameterisation of our adopted size distribution. For example, we assume that the size distribution varies only with $b_{\mathrm{C}}$, and that the other parameters such as $a_{\mathrm{0,}i}$ and $\sigma$ are fixed. Although~\citet{Weingartner:01} were able to fit extinction curves for a range of $R_{\mathrm{V}}$ values with these parameters fixed to the values used in this analysis, it is possible that these parameters will vary. For instance, if grain growth is occurring, then this will likely lead to an increase in $a_{\mathrm{0,}i}$, which would have an impact on the amplitude of the spinning dust emission. However, in this analysis we follow the example of~\citet{Weingartner:01} in assuming that these parameters are fixed, which enables us to compare our constrained value of $b_{\mathrm{C}}$ with the value obtained by~\citet{Li:01} for the diffuse ISM.

Additionally, varying the electric dipole moment of the grains will also affect the result. Increasing or decreasing this parameter will result in a corresponding increase or decrease in the spinning dust emission. However, as discussed in Section~\ref{Subsec:Model_Spin_Dust}, our adopted dipole moment is rather conservative, and any such increase in this value would result in an increase in the predicted spinning dust curves, which would produce a corresponding decrease in the constrained $b_{\mathrm{C}}$ values, therefore not affecting the results of the analysis.

Throughout this analysis we assume that all of the cm emission~($S_{1~\mathrm{cm}}^{\mathrm{observed}}$) is due to spinning dust emission. However, as mentioned in Section~\ref{Sec:Spinning_Dust}, it is possible that there may be a contribution from magnetic dipole emission. If this is the case, then this will decrease the contribution of spinning dust emission to the total cm emission, thereby further enhancing the observed deficit of very small grains. Therefore, ignoring magnetic dipole emission throughout this analysis is a conservative approach, and does not impact our results.

Finally, we stress that this analysis is the first attempt to use spinning dust emission to characterise the properties of interstellar dust grains, and even with these limitations, this work demonstrates the possibilities of this new method, and as such we hope that it will be included in future studies of interstellar dust.

%%%%%%% Conclusions %%%%%%%%%%%%%%%%%%%%%%%%%%%%%%%%%%%%%%%%%%%%%%%%

\section{Conclusions}
\label{Sec:Conclusions}

By combining 1~cm CARMA measurements, with far-IR \textit{Herschel} observations, we have constrained the abundance of the population of very small dust grains in a sample of Galactic cold cores. To do this we exploited the recent analysis by~\citet{Tibbs:15a} who used \textit{Herschel} data to derive $\bar{n}$$_{\mathrm{H}}$ and $\bar{G}$$_{\mathrm{0}}$ for a sample of 34 Galactic cores. These physical properties, in addition to the $T_{\mathrm{gas}}$ values estimated from CO observations~\citep{Wu:12}, have allowed us to substantially narrow the parameter space for modelling the spinning dust emission. By combining these constrained quantities with additional parameters~($x_{\mathrm{H}}$, $x_{\mathrm{C}}$, $y$, and $\mu$), we used \textsc{spdust} to model the spinning dust emission and estimate the predicted spinning dust emission in each of the cores. This analysis was repeated for four different values of the total C abundance in very small grains, and we found that for all of the cores, the best match between the theory and the data were for values of $b_{\mathrm{C}}$~$\le$~1$\times$10$^{-5}$. Such values of the very small dust grain abundance are lower than what is observed in the diffuse ISM~($b_{\mathrm{C}}$~$=$~6$\times$10$^{-5}$) suggestive of a deficit of very small grains in the cores. Based on timescale arguments we conclude that this deficit of very small grains is likely due to the process of grain growth via accretion and coagulation.

In conclusion, we acknowledge that there are limitations to this analysis, but nevertheless this work represents the first use of spinning dust emission to characterise the physical properties of interstellar dust grains, and we hope that this research avenue will be followed by other authors.

%%%%%%% Acknowledgments %%%%%%%%%%%%%%%%%%%%%%%%%%%%%%%%%%%%%%%%%%%%

\section*{Acknowledgments}

We thank the anonymous referee for providing detailed comments that have improved the content of this paper.
This work has been performed within the framework of a NASA/ADP ROSES-2009 grant, no. 09-ADP09-0059.

%%%%%%%%%%%%%%%  BIBLIOGRAPHY  %%%%%%%%%%%%%%%

%%%%%%%%%%%%%%%%%%%%%%%%%%%%%%%%%%%%%%%%%%%%%%%%%%%%%%%%%%%%%%%%%%%%

\bsp % ``This paper has been produced using the ...''

\label{lastpage}

\end{document}